\documentclass[a4paper]{aa}
\pdfoutput=1

\usepackage{graphicx}
\usepackage{natbib}
\usepackage{txfonts}
\usepackage{color}
\usepackage[usenames,dvipsnames]{xcolor}  % comment out for referee option
\usepackage{multirow}

\newcommand{\kms}{km\,s$^{-1}$}

\newcommand{\figps}[1]{\resizebox{\hsize}{!}{\rotatebox{0}{\includegraphics{#1}}}}

\newcommand{\fifps}[2]{\centering\resizebox{#1}{!}{\includegraphics{#2}}}
\newcommand{\firrps}[2]{\resizebox{#1}{!}{\rotatebox{270}{\includegraphics{#2}}}}

\newcommand{\beq}{\begin{equation}}
\newcommand{\eeq}{\end{equation}}

\begin{document}
\title{Doppler imaging of chemical spots on magnetic Ap/Bp stars}
\subtitle{Numerical tests and assessment of systematic errors}

\author{O.~Kochukhov}

\institute{
Department of Physics and Astronomy, Uppsala University, Box 516, SE-75120 Uppsala, Sweden\\\email{oleg.kochukhov@physics.uu.se}
}

\date{Received 22 September 2016 / Accepted 16 November 2016}

\titlerunning{Doppler imaging of Ap/Bp stars}
\authorrunning{O.~Kochukhov}

\abstract
%context
{
Doppler imaging (DI) is a powerful spectroscopic inversion technique that enables conversion of a line profile time series into a two-dimensional map of the stellar surface inhomogeneities. DI has been repeatedly applied to reconstruct chemical spot topologies of magnetic Ap/Bp stars with the goal of understanding variability of these objects and gaining an insight into the physical processes responsible for spot formation. 
}
%aims
{
In this paper we investigate the accuracy of chemical abundance DI and assess the impact of several different systematic errors on the reconstructed spot maps.
}
%methods
{
We simulate spectroscopic observational data for two different Fe spot distributions with a surface abundance contrast of 1.5~dex in the presence of a moderately strong dipolar magnetic field. We then reconstruct chemical maps using different sets of spectral lines and making different assumptions about line formation in the inversion calculations.
}
%results
{
Our numerical experiments demonstrate that a modern DI code successfully recovers the input chemical spot distributions comprised of multiple circular spots at different latitudes or an element overabundance belt at the magnetic equator. For the optimal reconstruction based on half a dozen spectral intervals the average reconstruction errors do not exceed $\sim$\,0.10 dex. The errors increase to about 0.15~dex when abundance distributions are recovered from a few and/or blended spectral lines. Ignoring a 2.5~kG dipolar magnetic field in chemical abundance DI leads to an average relative error of 0.2~dex and maximum errors of 0.3~dex. Similar errors are encountered if a DI inversion is carried out neglecting a non-uniform continuum brightness distribution and variation of the local atmospheric structure. None of the considered systematic effects leads to major spurious features in the recovered abundance maps. 
}
%conclusions
{
This series of numerical DI simulations proves that inversions based on 1--2 spectral lines with simplifying assumptions of the non-magnetic radiative transfer and a single model atmosphere  are generally reliable provided that the stellar magnetic field is not much stronger than 2--3~kG and the recovered spot map has a contrast of at least $\sim$\,0.3~dex. In the light of these findings we assess magnetic field strengths of several dozen Ap/Bp stars previously studied with DI methods, concluding that the vast majority of the published chemical spot maps are  unaffected by the systematic errors addressed in this paper.
}


\maketitle

\section{Introduction}
\label{intro}

Intermediate-mass chemically peculiar (CP) stars are A and B-type main sequence objects exhibiting unusual surface characteristics compared to normal stars of a similar spectral class. CP stars rotate slowly and show extreme chemical anomalies in their atmospheres, often corresponding to deviations from the solar chemical composition by several orders of magnitude for some heavy elements. This unusual surface chemistry is attributed to the processes of radiative levitation and gravitational settling, commonly referred to as atomic diffusion \citep{michaud:1970,michaud:2015}, operating in stably stratified stellar atmospheres.

A subgroup of CP stars, known as magnetic Ap/Bp or CP2 stars according to the classification by \citet{preston:1974}, possesses strong, stable, globally-organised magnetic fields on their surfaces. These fields, described to a first approximation by oblique dipolar geometries, are believed to be fossil remnants of the magnetic flux generated by dynamo or acquired from interstellar medium at an earlier evolutionary phase. The kG-strength fields further stabilise stellar atmospheres and facilitate diffusive separation of chemical elements; consequently magnetic Ap/Bp stars exhibit the most extreme chemical anomalies among CP stars. Moreover, due to an anisotropic and non-axisymmetric character of the global magnetic field topologies of Ap/Bp stars, the atomic diffusion process operates differently at different parts of the stellar surface, resulting in substantial horizontal and vertical chemical abundance gradients \citep{michaud:1981,babel:1991}.

Chemical inhomogeneities in the atmospheres of magnetic Ap/Bp stars have profound observational manifestations in the photometric, spectroscopic and spectrophotometric behaviour of these stars. All these different types of observables show stable, strictly repeating variation with the same period as the change of the disk-integrated magnetic field characteristics. This coherent variability is therefore commonly interpreted in the framework of the so-called oblique rotator model \citep{stibbs:1950}, postulating that the surface chemical and magnetic field distributions are static and the sole reason for the observed variability is rotational modulation of the aspect at which the surface structure is viewed by a distant observer.

Since the introduction of the oblique rotator model considerable effort was spent on deriving the geometry of chemical spots on Ap/Bp stars and understanding their relation to the surface magnetic field topology. The attempts to map the surface structure of magnetic CP stars dates back to the work by \citet{deutsch:1958,deutsch:1970}, who developed a technique of fitting phase curves of magnetic, equivalent width and radial velocity measurements with surface maps expanded in spherical harmonics. A more refined star-spot mapping approach, utilising the entire information content of stellar line profiles, was introduced by Khokhlova and collaborators \citep{khokhlova:1975,khokhlova:1976}. Their method, now commonly known as Doppler imaging (DI)\footnote{The term ``Doppler imaging'' was suggested later by \citet{vogt:1983a}.}, exploits spatial resolution of the stellar surface by the rotational Doppler effect. Provided the local (intrinsic) spectral line width is substantially smaller than the stellar projected rotational velocity, any inhomogeneity on the stellar surface yields a distortion in the disk-integrated line profile displaced from the line centre according to the local rotational Doppler shift and hence the spot longitude. This distortion moves across the line profile as the star rotates. Spots located at different latitudes exhibit different temporal variability patterns. For example, a spot near the stellar rotational equator produces a signature that is visible during half of the rotation period and moves quickly across the line profile. On the other hand, spots close to the rotational pole are visible during a larger fraction of the rotational cycle; the corresponding line profile distortions are constrained to the line centre.

Initial attempts to apply DI to line profile observations of Ap/Bp stars were based on the trial-and-error approach, often leading to subjective and non-unique solutions \citep{megessier:1979}. A breakthrough contribution was made by \citet{goncharskii:1977,goncharskii:1982}, who recasted the DI technique as a least-squares, regularised image reconstruction inverse problem (see \citet{kochukhov:2016} for a recent review). In this formulation a given chemical element map is specified on a discrete latitude-longitude surface grid and local abundances in each pixel of the map are iteratively adjusted by fitting model spectra to the observed spectral line profiles. The surface abundance distribution is further constrained by a regularisation function. This is a key mathematical ingredient of solving an ill-posed inverse problem. Regularisation ensures convergence to a unique solution and stability of the inversion results with respect to observational noise. 

The initial and most current applications of DI to chemical spot mapping on Ap/Bp stars relied on the Tikhonov regularisation method \citep{tikhonov:1977}. A smaller number of studies \citep[e.g.][]{hatzes:1989} used an alternative Maximum Entropy regularisation function. Although the external criteria imposed by the two regularisation methods are somewhat different (Tikhonov regularisation favours solutions with the smallest local gradients, Maximum Entropy favours solutions with the least deviation from a default value), it was demonstrated that DI maps obtained with alternative regularisation strategies are generally compatible with each other \citep{piskunov:1990a,strassmeier:1991} provided that they are based on the observational data with sufficiently high signal-to-noise ratio and a dense rotational phase coverage.

Early DI studies of Ap stars typically presented maps of a few chemical elements based on 1--2 isolated spectral lines for each element. Analytical functions were often employed to describe the local line profiles \citep[e.g.][]{khokhlova:1986,hatzes:1990,rice:1990}. As computational resources became more affordable, DI mapping could be extended to 5--10 chemical species, using multiple lines for a given element whenever possible and treating line blends with detailed radiative transfer calculations in realistic model atmospheres \citep[e.g.][]{kochukhov:2004e,luftinger:2010a,nesvacil:2012}. 

However, some important simplifications have lingered. Looking at the bulk of published DI maps of Ap/Bp stars one can conclude that the most common approach has been to derive chemical abundance distributions from a small number, sometimes just one or two, spectral lines, using a single model atmosphere and ignoring effects of magnetic field in calculation of the model spectra. Several recent studies took advantage of the magnetic DI \citep[MDI,][]{piskunov:2002a} extension of the indirect mapping technique to recover self-consistently magnetic field topologies and chemical spot distributions from the variation of Stokes parameter spectra \citep[e.g.][]{kochukhov:2015,silvester:2015,rusomarov:2016}. However, only a few stars were analysed with this sophisticated MDI method; most observational constraints on the chemical element distributions are still provided by non-magnetic DI investigations. Even fewer DI studies have addressed the problem of a feedback of chemical spots on the local atmospheric structure \citep{lehmann:2007,kochukhov:2012}\footnote{At the same time, a series of studies \citep{krticka:2007,krticka:2012,krticka:2015,shulyak:2010b} have modelled photometric and spectrophotometric variation of several Ap/Bp stars based on calculation of theoretical multi-component model atmospheres with the local abundances taken from the DI analyses of these stars.}.

The question of reliability of abundance DI inversions was repeatedly addressed by numerical tests in which known surface maps were reconstructed from simulated observations \citep{hatzes:1989,rice:1989,piskunov:1990b,rice:1991a,wehlau:1993,kochukhov:2002c}. These studies demonstrated that diverse, often very complicated, surface chemical over- and underabundance patterns (circular, square and complex letter-shaped spots, rings of different sizes) are recovered satisfactorily by the same inversion codes as applied to real stellar observations. Previous simulations also provided a comprehensive assessment of the sensitivity of DI maps to the value and uncertainty of key stellar parameters (projected rotational velocity and inclination angle) and to the quality (signal-to-noise ration, phase coverage, resolution) of observational data. However, several sources of potentially important systematic errors inherent to non-magnetic DI inversions have not been addressed before. The goal of this paper is therefore to present a new series of DI simulations and tests using the most sophisticated and physically realistic version of the indirect abundance mapping technique currently available. Based on these calculations, we quantify the loss of information and appearance of spurious structures in DI maps due to using few and/or blended spectral lines, neglecting the moderately strong magnetic field, and using a single average model atmosphere structure. A characterisation of these error sources enables an informed judgement on the level of reliability of several dozen abundance DI studies published over the past 30 years.

This paper is organised as follows. Sect.~\ref{codes} reviews different versions of Doppler imaging codes applied in previous studies of Ap/Bp stars and in this paper. Sect.~\ref{exp} describes the setup adopted for numerical DI experiments and discusses the results of the inversions using different line lists and different assumptions about the local line formation physics. In Sect.~\ref{disc} we present a literature survey of DI studies, assessing in light of our simulations the reliability of chemical spot maps published for 40 CP stars. The main conclusions of our study are summarised in Sect.~\ref{summary}.

\section{DI and MDI inversion codes}
\label{codes}

\subsection{{\sc Invers8}}

This is an older non-magnetic abundance DI code \citep{piskunov:1993} used in many previous DI studies of Ap/Bp stars reviewed in Sect.~\ref{disc}. The spectrum synthesis calculations by {\sc Invers8} are based on interpolation in the precomputed local line intensity tables, which are generated by approximating abundance variation of a single chemical element by scaling its line opacity. A single model atmosphere file is used and continuum intensity is considered to be independent of the local abundance. Inversions can incorporate several wavelength regions.

Despite relying on a single model atmosphere in its standard implementation, {\sc Invers8} can be easily modified to calculate the local line profile and continuum intensity tables using a grid of model atmospheres with varying abundance \citep{lehmann:2007}.

This code is not used in the numerical experiments discussed here.

\subsection{{\sc Invers11} and {\sc Invers12}}

These are newer non-magnetic abundance DI codes \citep{lueftinger:2003,kochukhov:2004e}. They are based on the same physical principles and assumptions as {\sc Invers8} but calculate the local line profiles on the fly and enable simultaneous reconstruction of several chemical maps from observations in multiple wavelength regions and from blended spectral features. Radiative transfer calculations in {\sc Invers12} are distributed over multiple processors with the Message Passing Interface (MPI) libraries.

These codes are not used in the numerical experiments presented in this paper.

\subsection{{\sc Invers10}}

This is the main magnetic DI code currently applied to Ap/Bp stars. The physical basis and numerical methods implemented in {\sc Invers10} were described in detail by \citet{piskunov:2002a}; the code was thoroughly tested by \citet{kochukhov:2002c}. The four Stokes parameter spectra computed with {\sc Invers10} agree with the calculations by other independent polarised radiative transfer codes \citep{wade:2001} and are considered as a benchmark for stellar Stokes parameter synthesis \citep{carroll:2008,deen:2013}.

{\sc Invers10} reconstructs simultaneously the three magnetic field vector components and an arbitrary number of chemical abundance maps from the Stokes $IQUV$ or Stokes $IV$ parameter spectra in multiple wavelength regions. The code can also be used for abundance mapping with Stokes $I$ alone, using a prescribed, constant magnetic field topology or ignoring the field altogether. {\sc Invers10} uses a single model atmosphere for numerical polarised radiative transfer calculations, which are performed on the fly and parallelised with MPI. Chemical abundance variations are approximated by scaling the central opacity of relevant spectral lines. Thus, no influence of chemical spots on the continuum intensity and on the atmospheric structure is taken into account. {\sc Invers10} also includes an option to automatically adjust oscillator strengths of selected spectral lines to mitigate the line-to-line mean equivalent width scatter due to errors in the input atomic data and/or neglect of some physical effects (non-local thermodynamical equilibrium, vertical chemical stratification).

All test inversions presented in this study were performed with {\sc Invers10}.

\subsection{{\sc Invers13}}

This MDI code represents a modification of {\sc Invers10} intended primarily for mapping magnetic fields and temperature spots on active cool stars \citep{kochukhov:2013}. {\sc Invers13} uses a grid of model atmospheres to calculate the local continuum intensity and Stokes parameter spectra according to the local temperature. It also incorporates an advanced equation of state solver suitable for treating molecular equilibrium at low temperatures \citep{piskunov:2016}. Thus, it can equally successfully model atomic and molecular lines. In the context of abundance mapping of Ap/Bp stars, a self-consistent treatment of the line formation and a model atmosphere structure allows one to use {\sc Invers13} for recovering abundance maps fully accounting for the influence of chemical inhomogeneities on the local atmosphere \citep{kochukhov:2012,oksala:2015}. In this case individual models in the input atmospheric grid, usually calculated with the {\sc LLmodels} code \citep{shulyak:2004}, have different chemical abundance rather than different temperature. To this end, {\sc Invers13} implements the most detailed and sophisticated simultaneous magnetic and chemical abundance inversion methodology. However, its drawback compared to {\sc Invers10} is the ability to model only one chemical element at a time.

In this paper we used {\sc Invers13} to generate simulated observations in Sect.~\ref{i13test}.

\subsection{{\sc InversLSD}}

This MDI code, developed by \citet{kochukhov:2014}, is designed for performing star spot and magnetic field inversions using least-squares deconvolved \citep[LSD,][]{donati:1997,kochukhov:2010a} Stokes parameter profiles. The spectrum synthesis calculations by {\sc InversLSD} are based on interpolation in the precomputed local LSD Stokes parameter profile tables. The latter can be generated either using simplified analytical solutions (weak-field approximation, Unno-Rachkovsky solution) of the polarised radiative transfer equation \citep[e.g.][]{kochukhov:2016a} or by applying the LSD line-averaging procedure to the  detailed local polarised radiative transfer calculations of the entire stellar spectrum \citep{kochukhov:2014,rosen:2015}. In addition to mapping the stellar magnetic field topology, {\sc InversLSD} allows one to recover a single scalar map (continuum brightness, temperature, abundance of a single chemical element).

This magnetic inversion code is not used in the numerical experiments presented in this paper.

\section{Numerical experiments}
\label{exp}

\subsection{Simulation setup}

For the numerical tests of abundance DI described below we adopted the following stellar parameters: effective temperature $T_{\rm eff}=8000$~K, surface gravity $\log g=4.0$, projected rotational velocity $v_{\rm e}\sin i=40$~\kms, inclination angle $i=60\degr$. The adopted $v_{\rm e}\sin i$ and $i$ are in the middle of, a relatively wide, range optimal for DI. The chosen atmospheric parameters represent cool Ap stars to which multi-element abundance mapping has been applied in a number of recent studies \citep[e.g.][]{kochukhov:2004e,luftinger:2010,nesvacil:2012}.

The corresponding stellar model atmosphere employed in spectrum synthesis was calculated with the {\sc LLmodels} code \citep{shulyak:2004} for the iron abundance $\log N_{\rm Fe}/N_{\rm tot}=-4.0$ and solar abundance of other elements. A grid of {\sc LLmodels} atmospheres with $\log N_{\rm Fe}/N_{\rm tot}$ spanning a range from $-2.5$ to $-4.0$ with a 0.25~dex step was generated for the multi-component atmosphere inversion test described in Sect.~\ref{i13test}.

\begin{figure*}[!th]
\centering
\firrps{15cm}{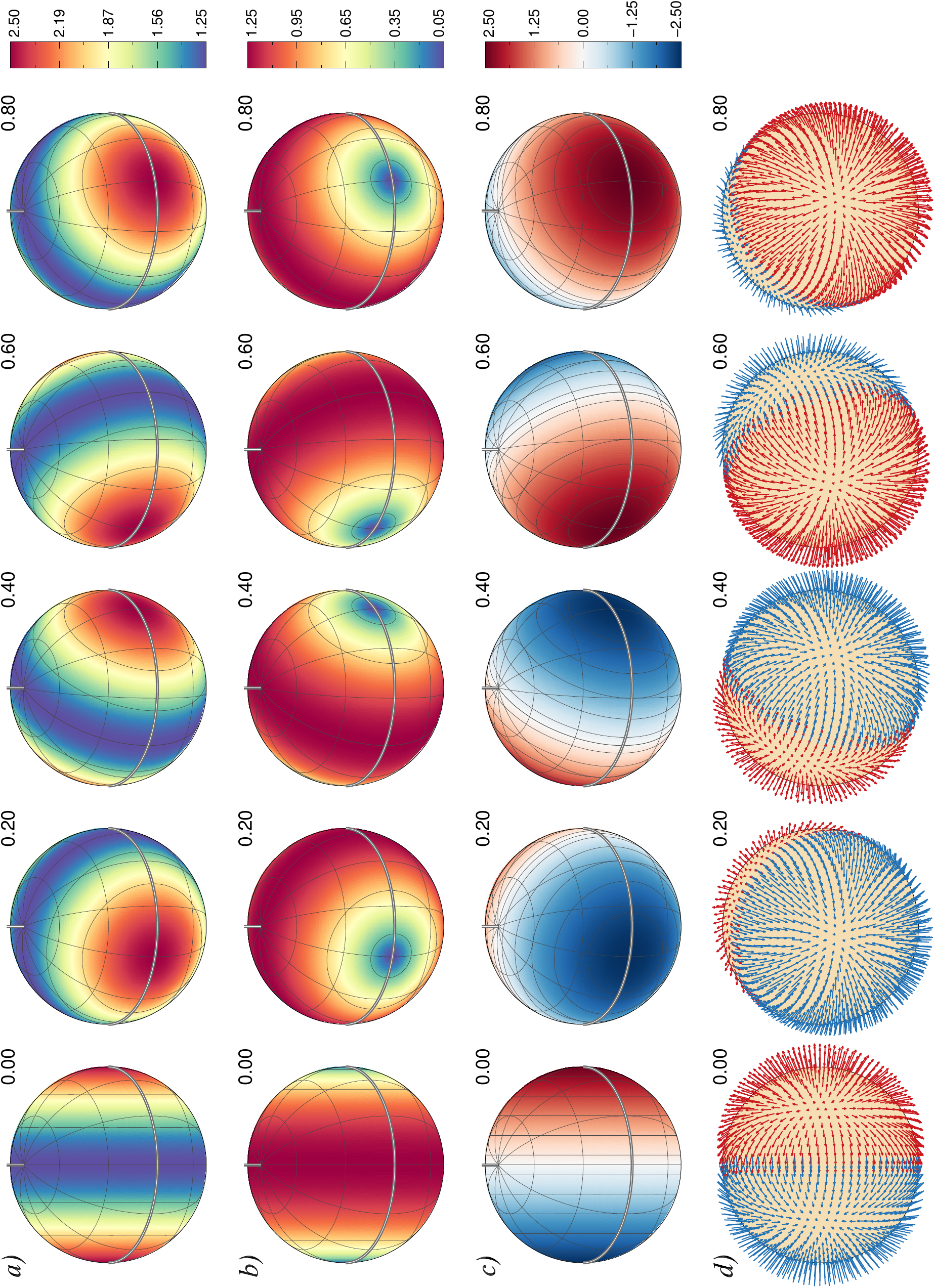}
\caption{Surface magnetic field topology adopted for numerical tests of DI. The star is shown at five rotation phases, which are indicated above the spherical plots, and an inclination angle of $i=60\degr$. The spherical plots show the maps of {\bf a)} field modulus, {\bf b)} horizontal field, {\bf c)} radial field, and {\bf d)} field orientation. The contours over spherical maps are plotted with a 0.5~kG step. The thick line and the vertical bar indicate positions of the rotational equator and the pole, respectively. The colour bars give the field strength in kG. The two different colours in the field orientation map correspond to the field vectors directed outwards (red) and inwards (blue).}
\label{fig:mag}
\end{figure*}

The star was assumed to have a dipolar magnetic field topology with a polar field strength of 2.5~kG and 90\degr\ inclination of the magnetic field axis relative to the rotational axis. This surface magnetic field configuration is illustrated in Fig.~\ref{fig:mag}.

Two different Fe surface abundance distributions were considered in the numerical experiments. The first comprised four circular overabundance spots, with $\log N_{\rm Fe}/N_{\rm tot}=-2.5$ in the spot centres and $\log N_{\rm Fe}/N_{\rm tot}=-4.0$ in the background. The spots were placed at the latitudes of $-30$, 0, 30, and 60\degr\ and spaced equidistantly in longitude. We adopted the inner spot radius of $r_{\rm in}=15\degr$ and the outer radius of $r_{\rm out}=30\degr$. The element abundance was linearly interpolated between $r_{\rm in}$ and $r_{\rm out}$, resulting in smooth spot edges. 

This multiple spot configuration is more complex than the 3-spot abundance map previously considered in the numerical tests of {\sc Invers10} \citep{kochukhov:2002c}. In particular, it includes a low-latitude feature which is challenging to recover since it is visible close to the stellar limb in a narrow range of rotational phases. It should also be noted that this four-spot configuration is more structured than the typical chemical abundance distributions found for magnetic Ap/Bp stars: for the majority of stars single or double-spot distributions are prevalent resulting in a single or double-wave photometric variability \citep[e.g.][]{bernhard:2015}.

The second chemical abundance distribution considered for DI experiments was represented by a ring of Fe overabundance ($\log N_{\rm Fe}/N_{\rm tot}=-2.5$) on a uniform background ($\log N_{\rm Fe}/N_{\rm tot}=-4.0$). Reckoning from the symmetry axis, the ring had $r_{\rm in}=85\degr$ and $r_{\rm out}=70\degr$ (yielding the width of 10\degr\ at maximum abundance and 40\degr\ at the base) and was chosen to coincide with the dipolar field equator. This relatively unusual element distribution pattern has been observed for certain elements in a few Ap stars \citep{rice:1997,kochukhov:2004e} but is predicted to be the most common non-uniform chemical spot distribution according to the theoretical atomic diffusion calculations in the presence of a magnetic field \citep{alecian:1981,leblanc:2009,alecian:2010,alecian:2015}.

In one test, described in Sect.~\ref{2lines}, we combined the two maps in a multi-element abundance mapping experiment. In that case the 4-spot distribution was adopted for Fe and the ring-like abundance configuration was used for Cr.

For most of the tests described below the DI calculations were based on seven \ion{Fe}{i} spectral lines distributed over 6\footnote{The lines \ion{Fe}{i} 5005.7 and 5006.1~\AA\ are blended for the adopted $v_{\rm e}\sin i$ of 40~\kms.} spectral intervals in the 5006--5397~\AA\ wavelength range. This group of neutral Fe lines was previously employed by \citet{luftinger:2010} in their DI study of HD\,24712. For the experiments in Sect.~\ref{2lines} the spectra were calculated for a blend of \ion{Fe}{ii} 6147.7, 6149.2~\AA\ and \ion{Cr}{ii} 6146.2~\AA\ lines used by \citet{kochukhov:2004e} in the DI study of HD\,83368. The main atomic parameters of the modelled transitions, including the effective Land\'e factors, are summarised in Table~\ref{tbl:lines}. This information was obtained from the {\sc VALD} database \citep{kupka:1999}.

\begin{table}[!th]
\centering
\caption{Spectral lines used in numerical tests of abundance DI. 
\label{tbl:lines}}
\begin{tabular}{lclrc}
\hline\hline
Ion & $\lambda$ (\AA) & $E_{\rm lo}$ (ev) & $\log gf$ & $\bar{g}$ \\
\hline
\ion{Fe}{i} & 5005.7122 &  3.884 & $0.029$ & 1.335 \\
\ion{Fe}{i} & 5006.1186 &  2.833 & $-0.638$ & 1.545 \\
\ion{Fe}{i} & 5242.4907 &  3.634 & $-0.967$ & 0.985 \\
\ion{Fe}{i} & 5263.3062 &  3.266 & $-0.879$ & 1.500 \\
\ion{Fe}{i} & 5367.4659 &  4.415 & $0.443$ & 0.915 \\
\ion{Fe}{i} & 5383.3685 &  4.313 & $0.645$ & 1.115 \\
\ion{Fe}{i} & 5397.1279 &  0.915 & $-1.993$ & 1.425 \\
\hline
\ion{Cr}{ii} & 6146.1788 &  4.756 & $-2.892$ & 1.515 \\
\ion{Fe}{ii} & 6147.7341 &  3.889 & $-2.827$ & 0.825 \\
\ion{Fe}{ii} & 6149.2459 &  3.889 & $-2.720$ & 1.350 \\
\hline
\end{tabular}
\tablefoot{The columns give the ion name, central wavelength, lower energy level, oscillator strength and the effective Land\'e factor.}
\end{table}

The simulated observational data were produced for 20 equidistant rotational phases. Spectra were convolved with a Gaussian instrumental profile corresponding to the resolution of $\lambda/\Delta\lambda=110\,000$ and were sampled with a step of $\Delta\lambda=0.02$~\AA. A normally distributed random noise component with $\sigma=1.9\times10^{-3}$ was added to each spectrum in the inversions with seven \ion{Fe}{i} lines. In the inversions with the two \ion{Fe}{ii} lines and the blended \ion{Cr}{ii} line we adopted $\sigma=10^{-3}$ to roughly offset an increase of random noise due to a smaller number of wavelength points in the simulated spectra. The adopted spectral resolution corresponds to the one provided by the HARPS-type spectrograph \citep{mayor:2003}. The assumed signal-to-noise ratio and phase coverage are realistic compared to recent observations of bright Ap stars \citep[e.g.][]{silvester:2012,wade:2016}.

Chemical abundance maps were recovered using a 1876-element surface grid with variable number of zones in 38 latitude belts (see Fig.~5 in \citet{piskunov:2002a} for an example of such a surface grid). All inversions started from a homogeneous element distribution with $\log N_{\rm el}/N_{\rm tot}=-4.0$.

\subsection{Choice of regularisation parameter}

A considerate choice of regularisation parameter is essential for a successful Doppler imaging inversion. A too strong regularisation will smooth out real features in a stellar surface map and will yield a poor fit to observations. Conversely, a too weak regularisation might result in a lot of unreliable high-contrast surface details. The usual approach to finding an optimal regularisation parameter ($\Lambda$) is to ensure a balance between reaching the lowest possible chi-square ($\chi^2$) of the fit to observations on the one hand and limiting unnecessary high-contrast details in the surface maps on the other hand. Our extensive experience with different DI problems suggests that an optimal regularisation parameter is such that i) no significant improvement of the fit quality is possible to achieve by decreasing $\Lambda$ and ii) the regularisation functional (i.e. regularisation function times the regularisation parameter) contribution to the total discrepancy function is a factor of 2--3 below the corresponding $\chi^2$ contribution.

\begin{figure}[!th]
\centering
\figps{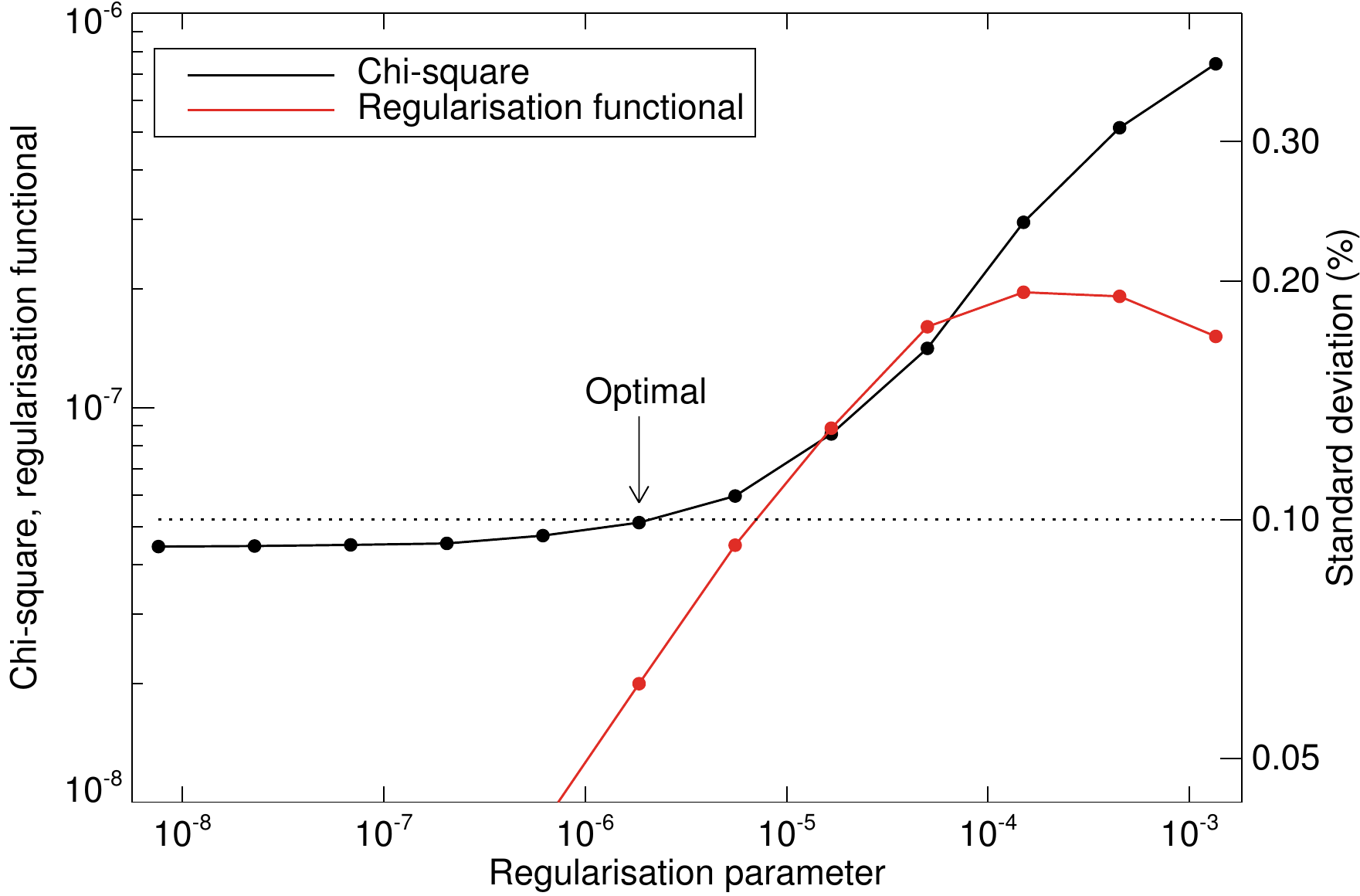}
\caption{Illustration of the regularisation parameter determination for the inversion with two \ion{Fe}{ii} lines. A measure of $\chi^2$ (\textit{solid black line)} and the regularisation functional (\textit{solid red line}) are plotted versus the Tikhonov regularisation parameter. The dotted line shows the noise level of the simulated data set. The optimal regularisation parameter is indicated with an arrow.}
\label{fig:regul}
\end{figure}

Figure~\ref{fig:regul} presents a validation of this empirical prescription for determining the Tikhonov regularisation parameter. This figure shows the $\chi^2$ and regularisation functional values for a series of 12 DI inversions (corresponding to the two \ion{Fe}{ii} line inversion test discussed in Sect.~\ref{2lines}) starting with a large $\Lambda$ and decreasing this parameter by a factor of 3 between consecutive inversions. The resulting dependence of $\chi^2$ on $\Lambda$ is characteristic of an inverse problem regularised with the Tikhonov method. Initially, the fit to observations is poor for large $\Lambda$; it rapidly improves as $\Lambda$ is let to decrease. At this stage regularisation functional is closely coupled and is comparable in magnitude to $\chi^2$. Then there is a break in the $\chi^2$ curve after which the fit quality is improving very slowly with decreasing $\Lambda$. At this point the regularisation functional decouples from the chi-square. The standard deviation of the fit reaches the nominal value of $10^{-3}$ when the regularisation functional is 2.6 times smaller than the chi-square.

This procedure of determining the Tikhonov regularisation parameter is closely related to the so-called L-curve method \citep{hansen:2000}, which looks for the point of maximum curvature in the $\log \Lambda$-$\log \chi^2$ plot. In practice, some uncertainty in determining $\Lambda$ is inevitable since the optimal fit quality is not known a priori and the $\chi^2$ variation with $\Lambda$ might not have a well defined break. Nevertheless, the usual uncertainty in $\Lambda$ is no more than a factor of 2--3; variation of the regularisation strength within these limits leads to a small and completely predictable modification of the resulting DI maps.

All DI calculations discussed in this paper were carried out with the step-wise regularisation parameter reduction, yielding 10--12 maps for different $\Lambda$ values. The maps presented below correspond to the regularisation parameter chosen in the reference inversions (Sects.~\ref{refs} and \ref{refc}) following the procedure described above. The maps presented for other inversions correspond to the same regularisation parameter as for the reference maps.

\begin{table*}[!th]
%\centering
\caption{Numerical tests of abundance DI inversions. 
\label{tbl:tests}}
\begin{tabular}{ccccccc}
\hline\hline
Abundance map & Magnetic field & Line list & $\sigma/\sigma_0$ &  $\langle\delta\varepsilon\rangle$ (dex) & $\langle|\delta\varepsilon|\rangle$ (dex) & $|\delta\varepsilon|_{75}$ (dex) \\
\hline 
\multicolumn{7}{c}{\textit{Reference inversions}}\\
4 spots & included & 7 \ion{Fe}{i} & 0.97 & 0.01 & 0.09, 0.07 & 0.12, 0.11 \\
ring & included & 7 \ion{Fe}{i} & 0.97 & 0.00 & 0.06, 0.06 & 0.08, 0.08 \\
\multicolumn{7}{c}{\textit{Reduced line list and blended lines}}\\
4 spots & included & 2 \ion{Fe}{ii} & 0.99 & 0.02 & 0.11, 0.10 & 0.15, 0.13 \\
4 spots (Fe) & included & 2 \ion{Fe}{ii} + 1 \ion{Cr}{ii} & 1.07 & 0.01 & 0.13, 0.11 & 0.16, 0.15 \\
ring (Cr) & included & 2 \ion{Fe}{ii} + 1 \ion{Cr}{ii} & 1.07 & 0.00 & 0.08, 0.08 & 0.11, 0.11 \\
\multicolumn{7}{c}{\textit{Ignoring magnetic field}}\\
4 spots & ignored & 7 \ion{Fe}{i} & 1.11 & 0.32 & 0.22, 0.19 & 0.33, 0.28 \\
ring & ignored & 7 \ion{Fe}{i} & 1.22 & 0.31 & 0.22, 0.19 & 0.32, 0.28 \\ 
uniform & ignored & 7 \ion{Fe}{i} & 1.09 & 0.34 & 0.13, 0.13 & 0.18, 0.19 \\
\multicolumn{7}{c}{\textit{Effect of local atmospheric structure}}\\
4 spots & included & 7 \ion{Fe}{i} & 1.25 & 0.05 & 0.21, 0.19 & 0.29, 0.27 \\
ring & included & 7 \ion{Fe}{i} & 2.16 & 0.01 & 0.21, 0.21 & 0.29, 0.29 \\
\hline
\end{tabular}
\tablefoot{The columns 1--3 describe the surface abundance distribution adopted for the test, indicate the treatment of magnetic field in chemical abundance DI, and provide information on the spectral line list used for the inversion. Column 4 lists the ratio of the final standard deviation of the model fit to the amplitude of the random noise in simulated data. Column 5 gives the mean offset between the true and reconstructed abundance map for latitudes $\ge$\,$-30\degr$; column 6 lists the mean absolute error of the reconstructed horizontal abundance structure for latitudes $\ge$\,$-30\degr$ and $\ge$\,$0\degr$; column 7 gives the 75th percentile of the abundance difference map for latitudes $\ge$\,$-30\degr$ and $\ge$\,$0\degr$.}
\end{table*}

\subsection{Error analysis}

We characterise the errors of abundance DI inversions by examining the difference between the true and reconstructed element distribution maps. Several complementary characteristics of the difference map are considered here and summarised in Table~\ref{tbl:tests} for the different inversions discussed below. First, we determined the average offset between the input and reconstructed map. This offset is of no concern for DI mapping since the primary goal of this technique is to determine the relative variation of an element concentration across the stellar surface, not the absolute abundance. The average offset was subtracted from the difference map before carrying out further statistical analysis. We then determined the mean absolute abundance difference. This gives an estimate of the average inversion errors. 

Of course, much higher discrepancies between the input and reconstructed maps are occasionally found locally at certain surface positions. The largest of these errors are irrelevant since they characterise a minor fraction of the stellar surface. Instead of discussing the maximum errors of individual surface elements we choose to consider discrepancies on the spatial scales comparable to the structures in adopted abundance maps. Given that the relative overabundance regions occupy from 27\% (4 spots) to 34\% (ring) of the stellar surface area, we adopt the 75th percentile of the unsigned difference map as a relevant measure of the maximum inversion errors.

In agreement with previous DI simulations \citep{berdyugina:1998a,piskunov:2002a}, most inversions reported here showed the largest errors and coherent artefacts in the sub-equatorial, poorly visible parts of the stellar surface. This is a natural consequence of the strong latitudinal dependence of the visibility and spectral contribution of individual surface zones. To roughly assess this latitudinal trend we report the average and maximum reconstruction errors separately for latitudes $\ge-30\degr$ (essentially the entire part of the stellar surface where a reasonable DI reconstruction can be expected for the adopted $i=60\degr$) and $\ge0\degr$ (the part of the surface with the best reconstruction quality).

\begin{figure*}[!th]
\centering
\firrps{15cm}{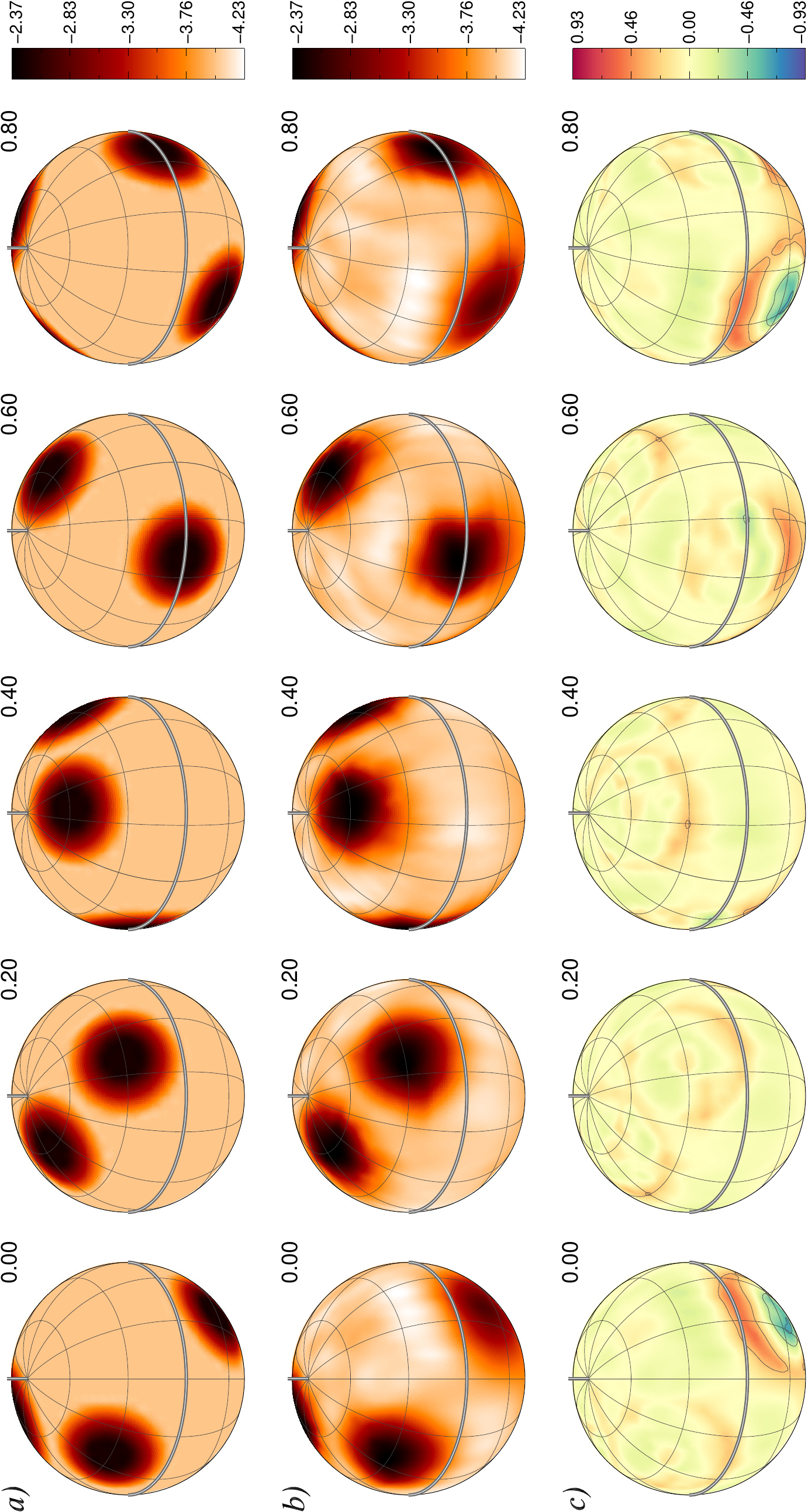}
\caption{Results of the Fe abundance inversion for the reference test case (4 circular spots, 7 \ion{Fe}{i} lines, including magnetic field). The star is shown at five rotation phases, which are indicated above the spherical plots, and an inclination angle of $i=60\degr$. The spherical plots show {\bf a)} the true abundance map, {\bf b)} reconstructed abundance map, {\bf c)} difference map. The contours over the difference map are plotted at $\pm0.3, 0.6, 0.9$ dex. The thick line and the vertical bar indicate positions of the rotational equator and the pole, respectively. The colour bars give the local abundance in $\log N_{\rm Fe}/N_{\rm tot}$ units.}
\label{fig:map1}
\end{figure*}

\begin{figure*}[!th]
\centering
\firrps{15cm}{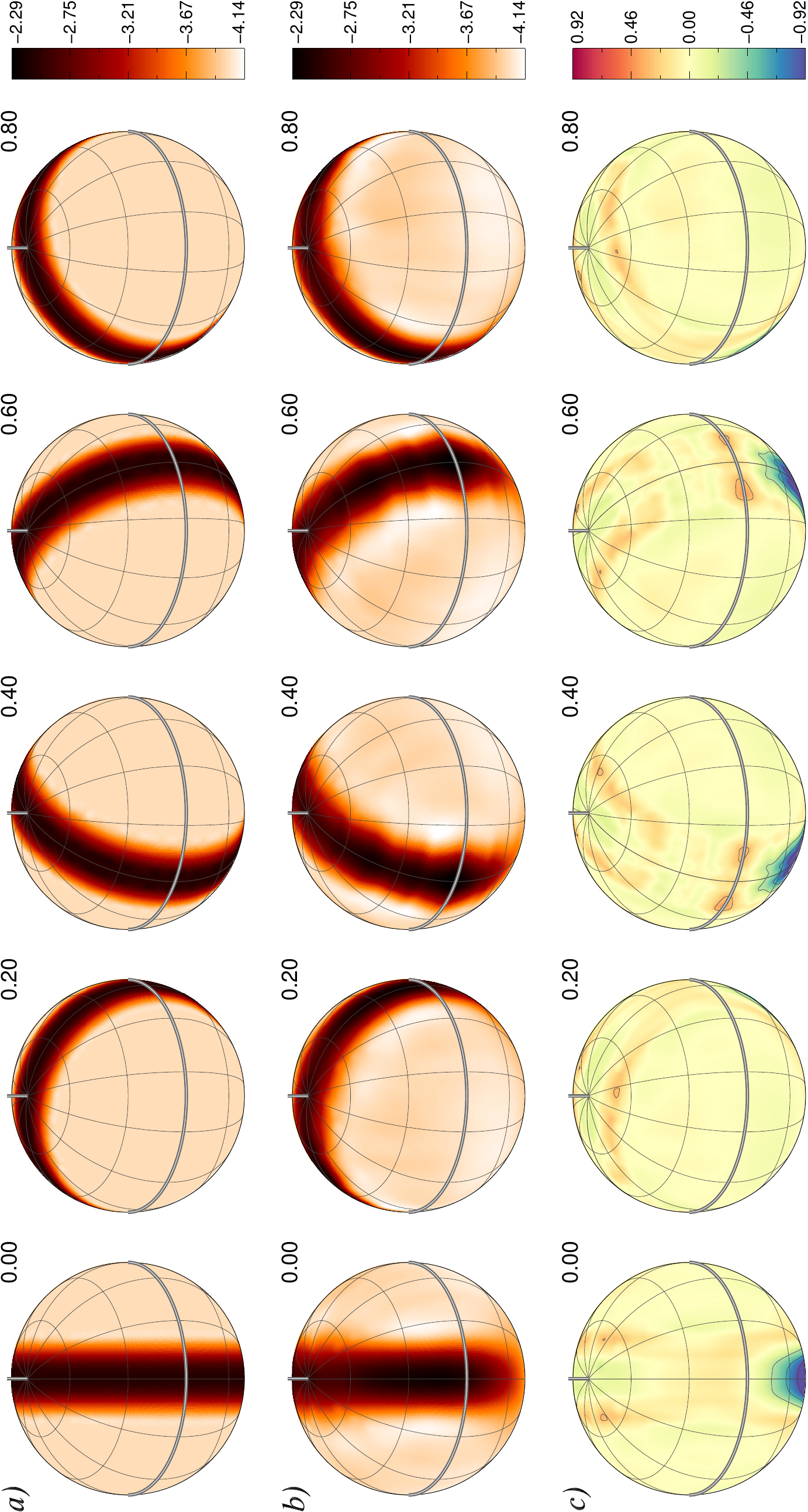}
\caption{Same as Fig.~\ref{fig:map1} but for the reconstruction of the overabundance ring.}
\label{fig:map2}
\end{figure*}

\begin{figure*}[!th]
\centering
\firrps{15cm}{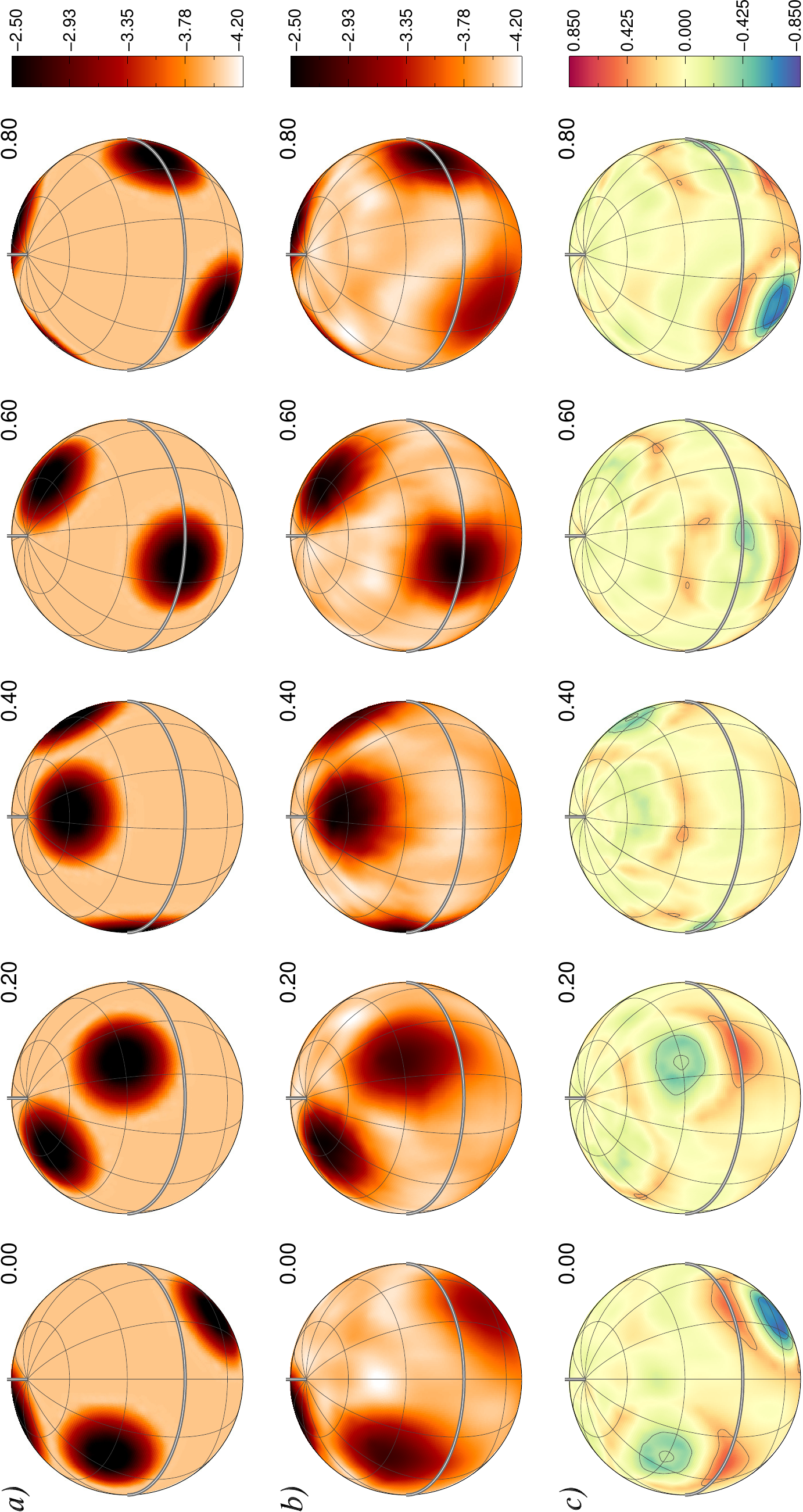}
\caption{Same as Fig.~\ref{fig:map1} but using two \ion{Fe}{ii} lines.}
\label{fig:map3}
\end{figure*}

\begin{figure*}[!th]
\centering
\firrps{15cm}{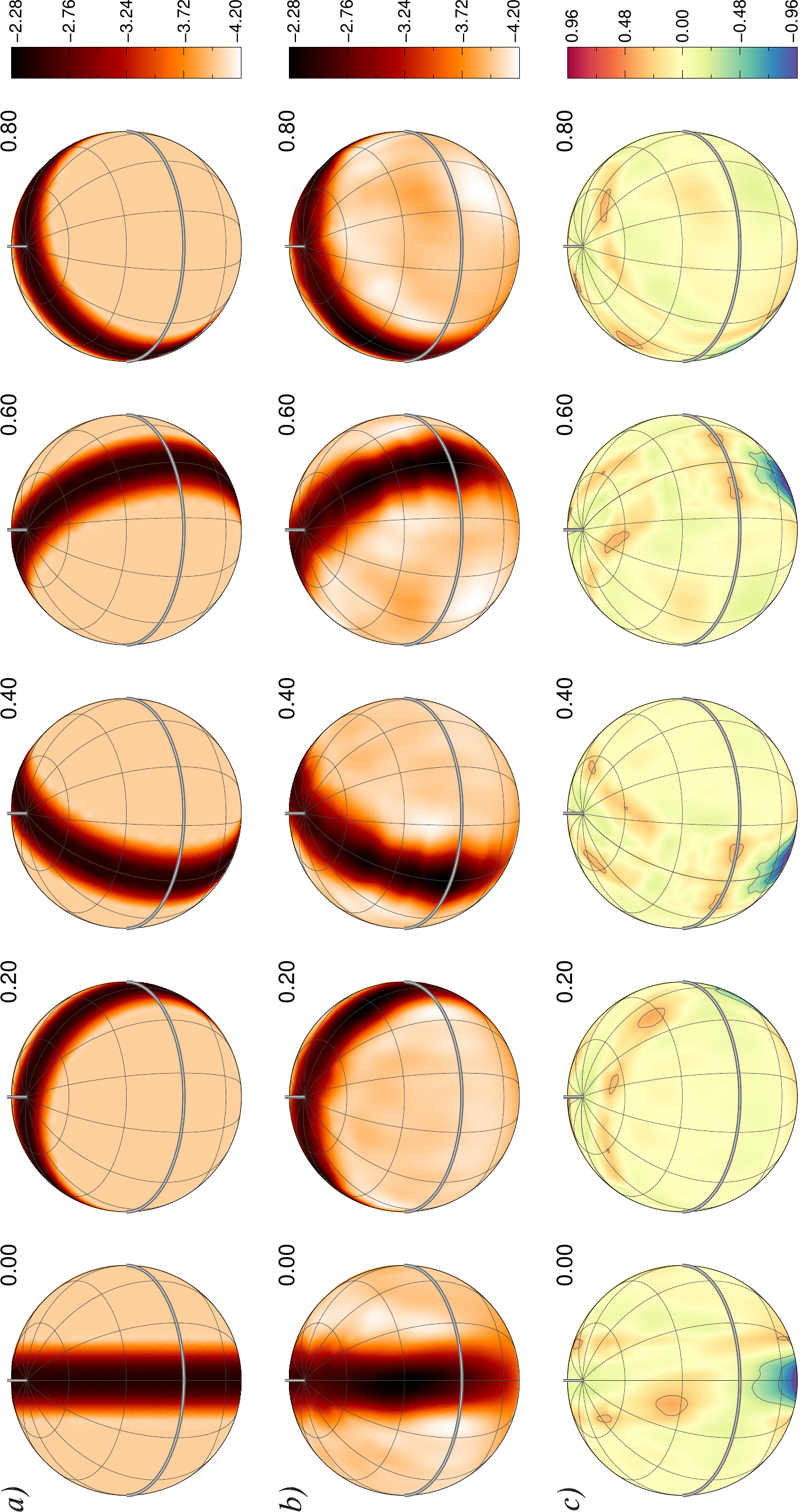}\vspace*{0.5cm}
\firrps{15cm}{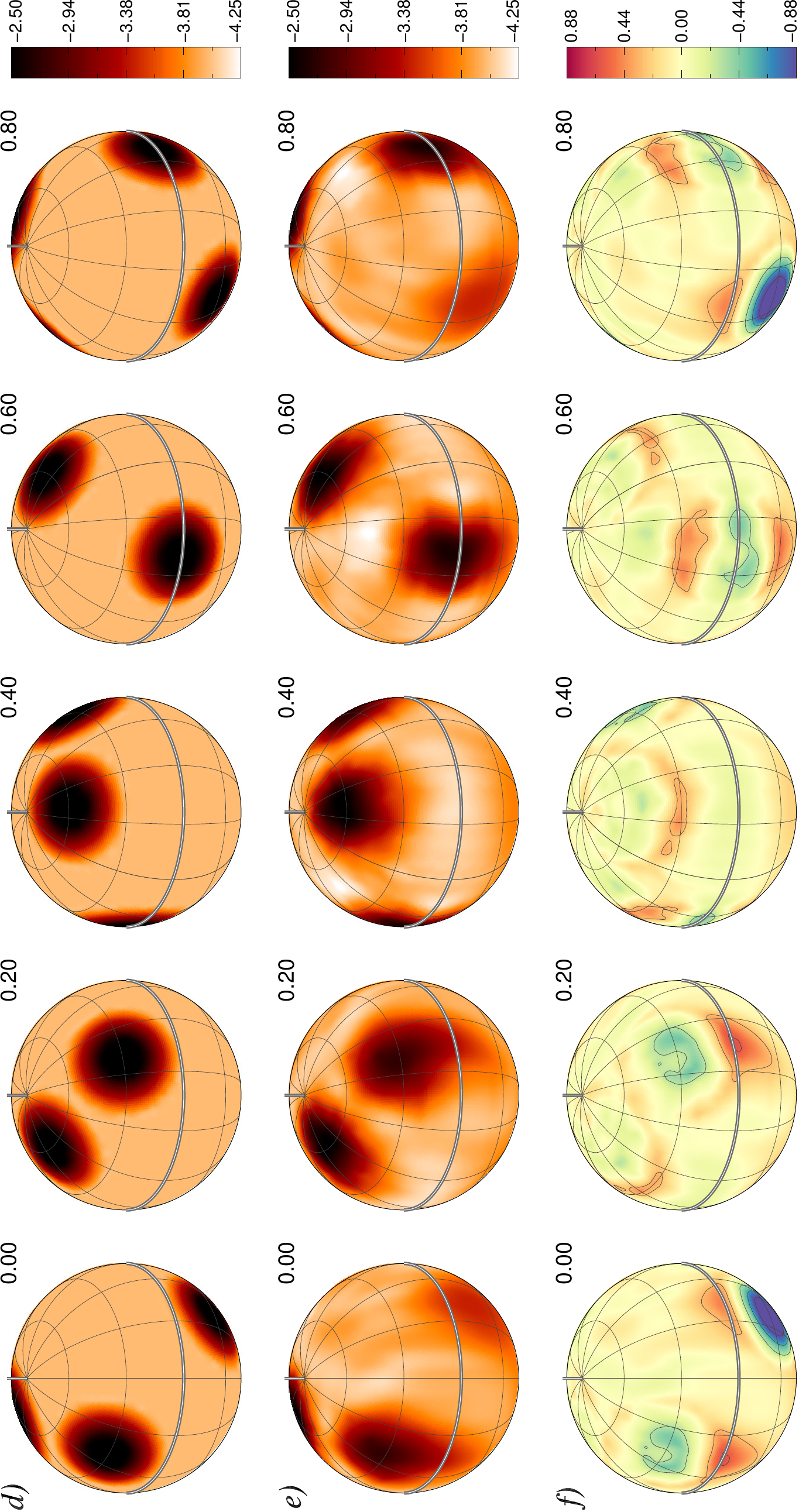}
\caption{Same as Fig.~\ref{fig:map1} but for the simultaneous reconstruction of Cr (panels {\bf a}-{\bf c}) and Fe (panels {\bf d}-{\bf f}) abundance maps using two \ion{Fe}{ii} lines and one blended line of \ion{Cr}{ii}.}
\label{fig:map4}
\end{figure*}

\begin{figure*}[!th]
\centering
\firrps{15cm}{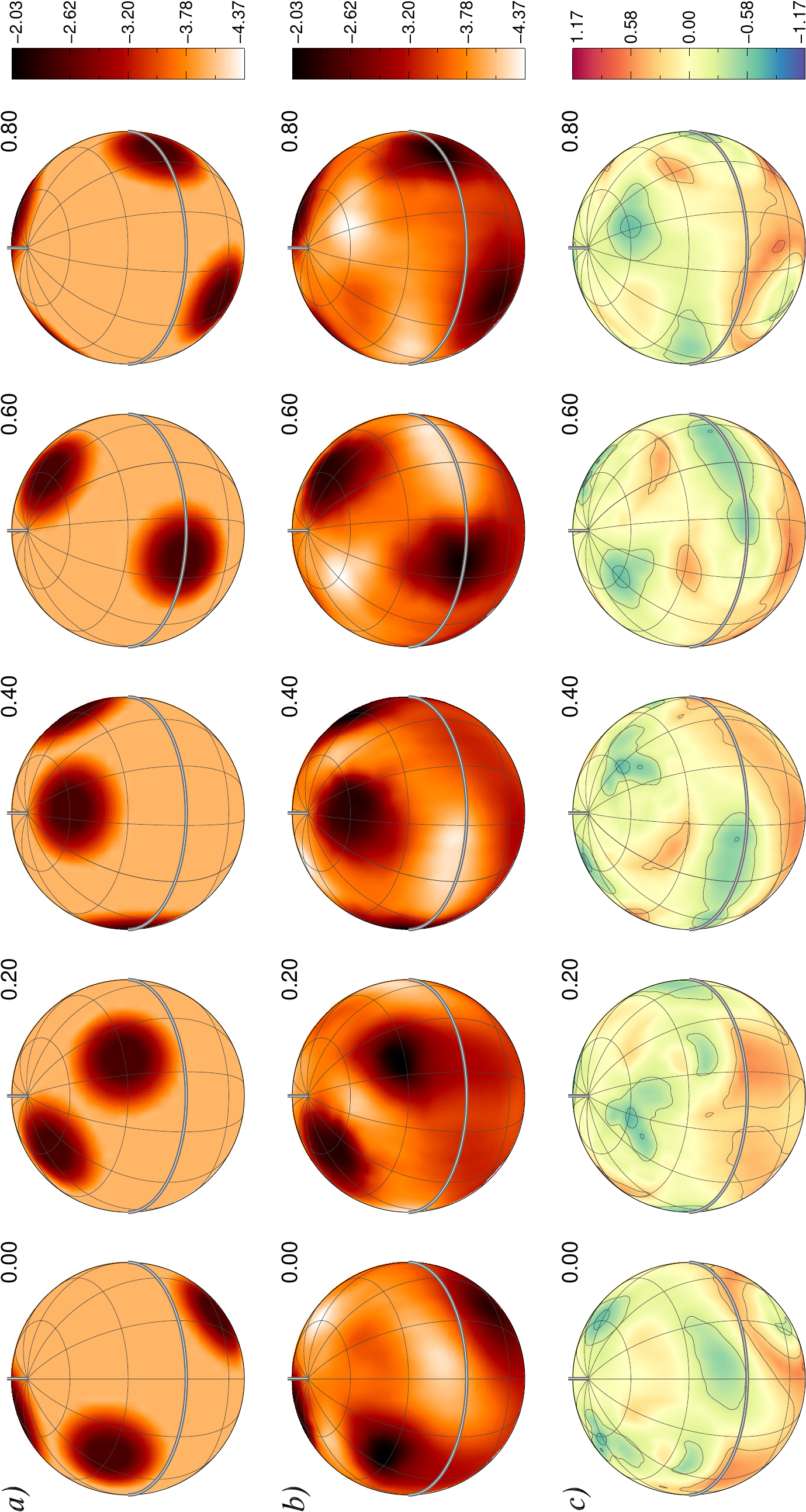}
\caption{Same as Fig.~\ref{fig:map1} but ignoring magnetic field. A constant offset is subtracted from the difference map shown in panel {\bf c)}.}
\label{fig:map5}
\end{figure*}

\begin{figure*}[!th]
\centering
\firrps{15cm}{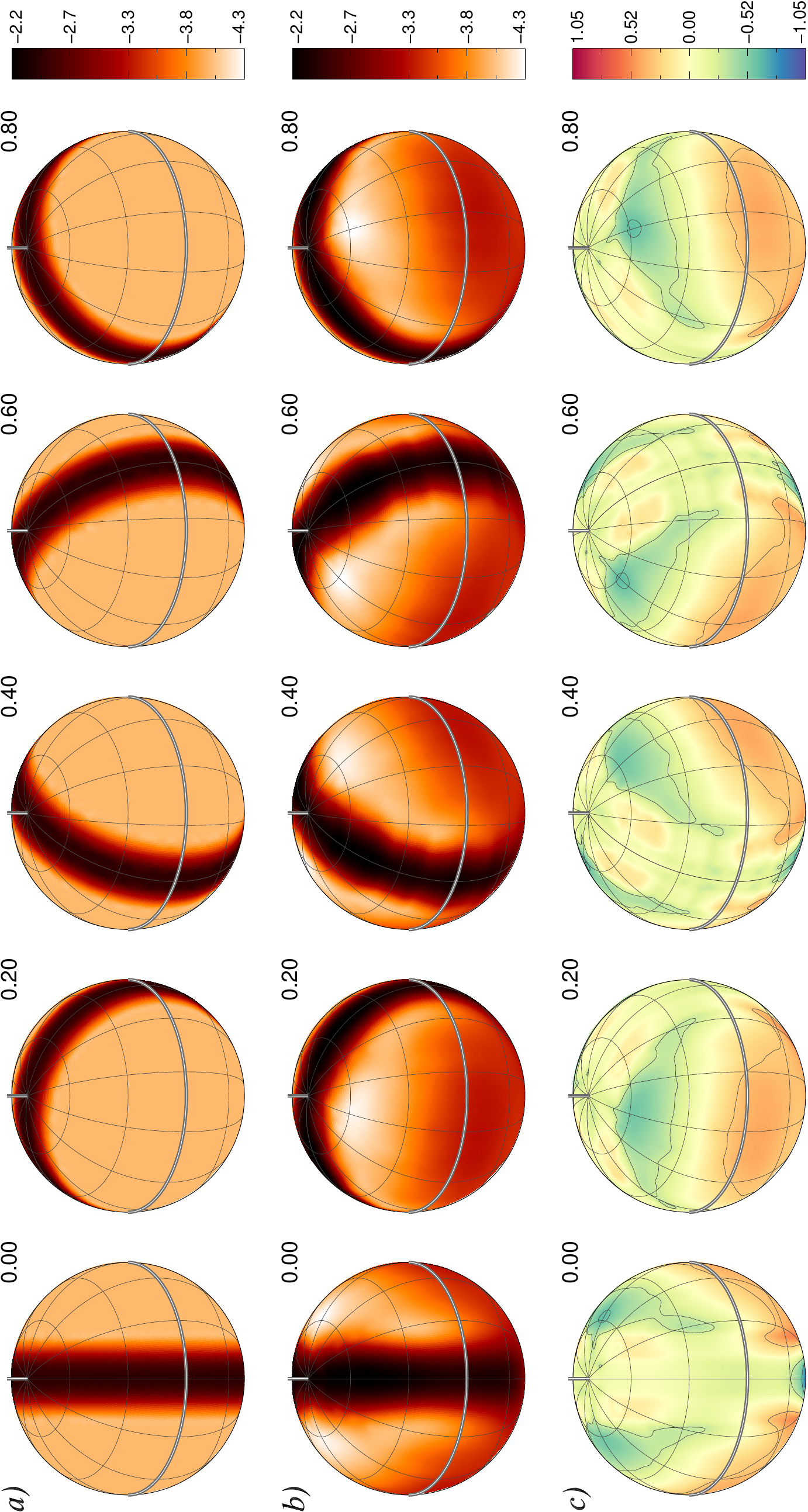}
\caption{Same as Fig.~\ref{fig:map2} but ignoring magnetic field. A constant offset is subtracted from the difference map shown in panel {\bf c)}.}
\label{fig:map6}
\end{figure*}

\begin{figure*}[!th]
\centering
\firrps{15cm}{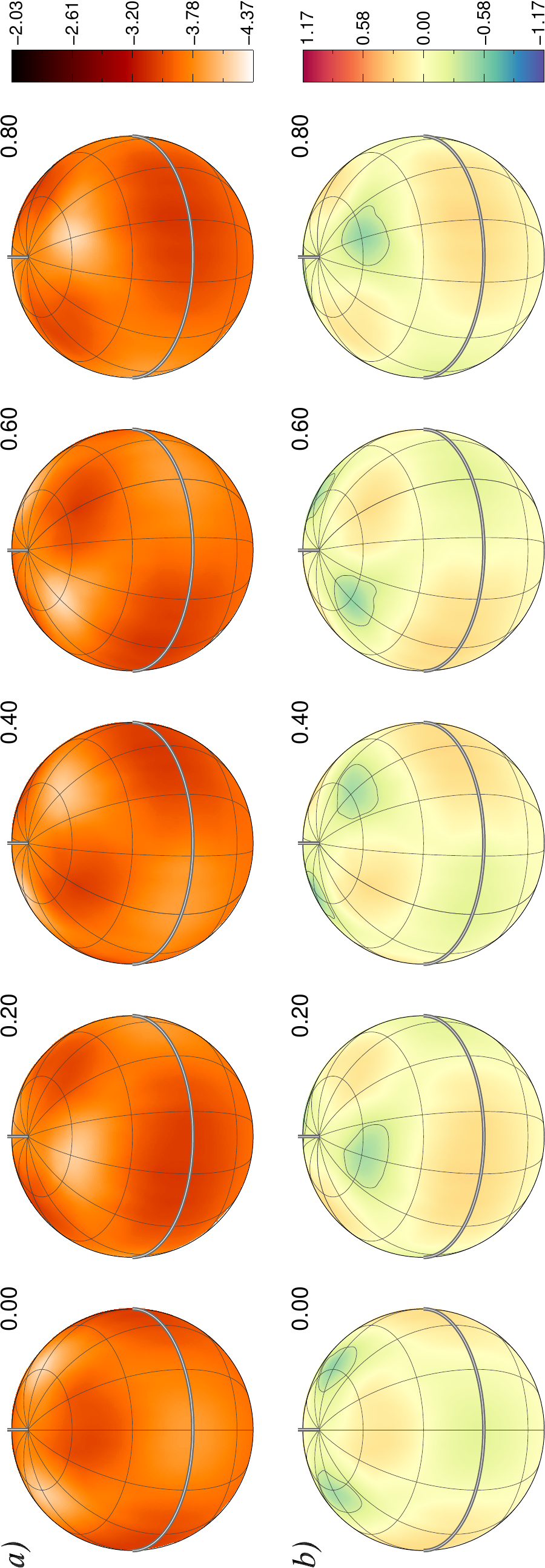}
\caption{Spurious chemical abundance distribution resulting from chemical mapping ignoring magnetic field. The true abundance map is a homogeneous distribution with $\log N_{\rm Fe}/N_{\rm tot}=-4.0$. Reconstructed Fe abundance map is shown in the top panel adopting the same abundance range as in Fig.~\ref{fig:map5}. The bottom panel shows the same distribution after subtracting a constant offset.}
\label{fig:map7}
\end{figure*}

\begin{figure*}[!th]
\centering
\firrps{15cm}{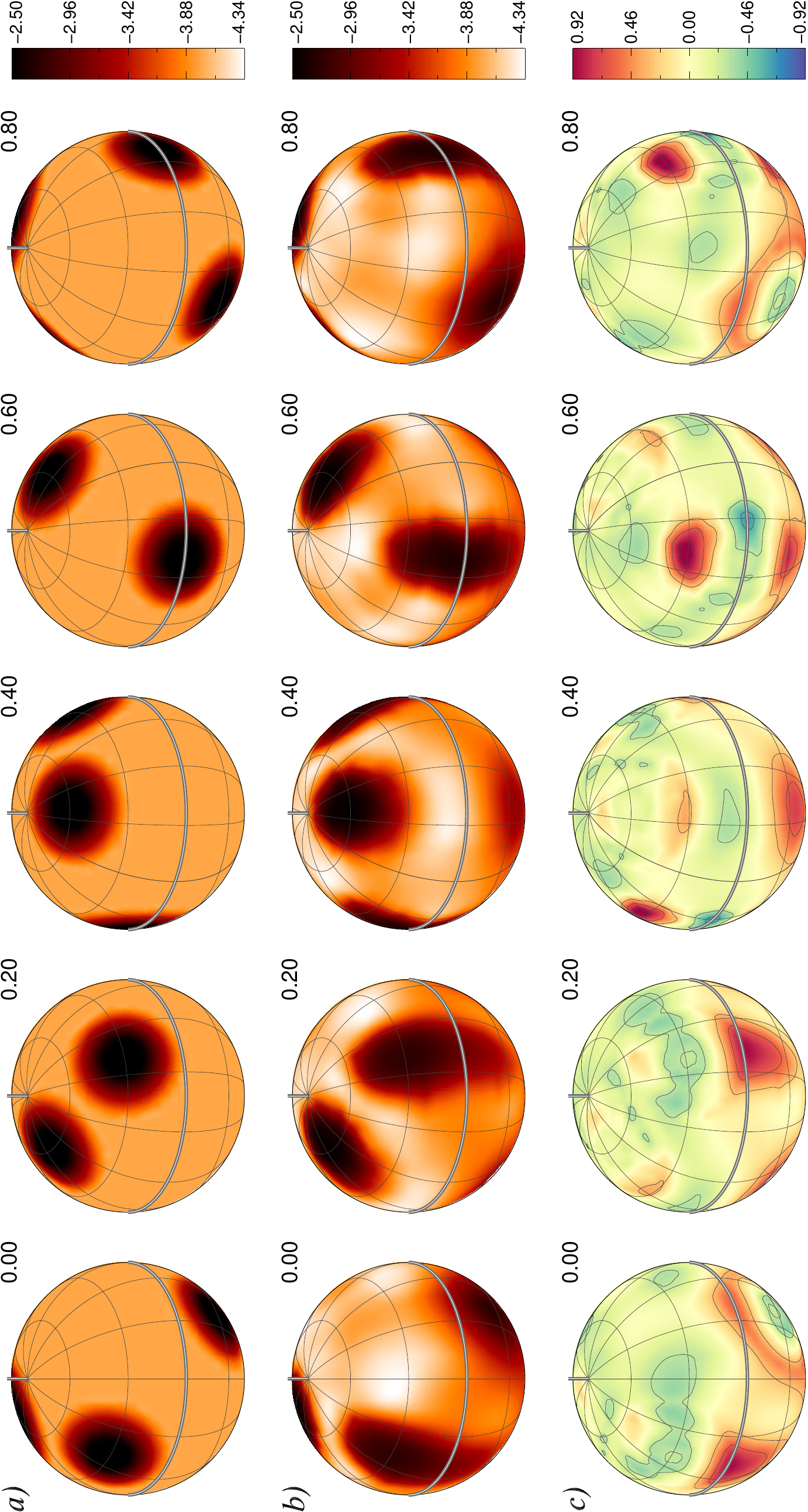}
\caption{Same as Fig.~\ref{fig:map1} but ignoring effect of local abundance variation on the model atmosphere structure. 
}
\label{fig:map8}
\end{figure*}

\begin{figure*}[!th]
\centering
\firrps{15cm}{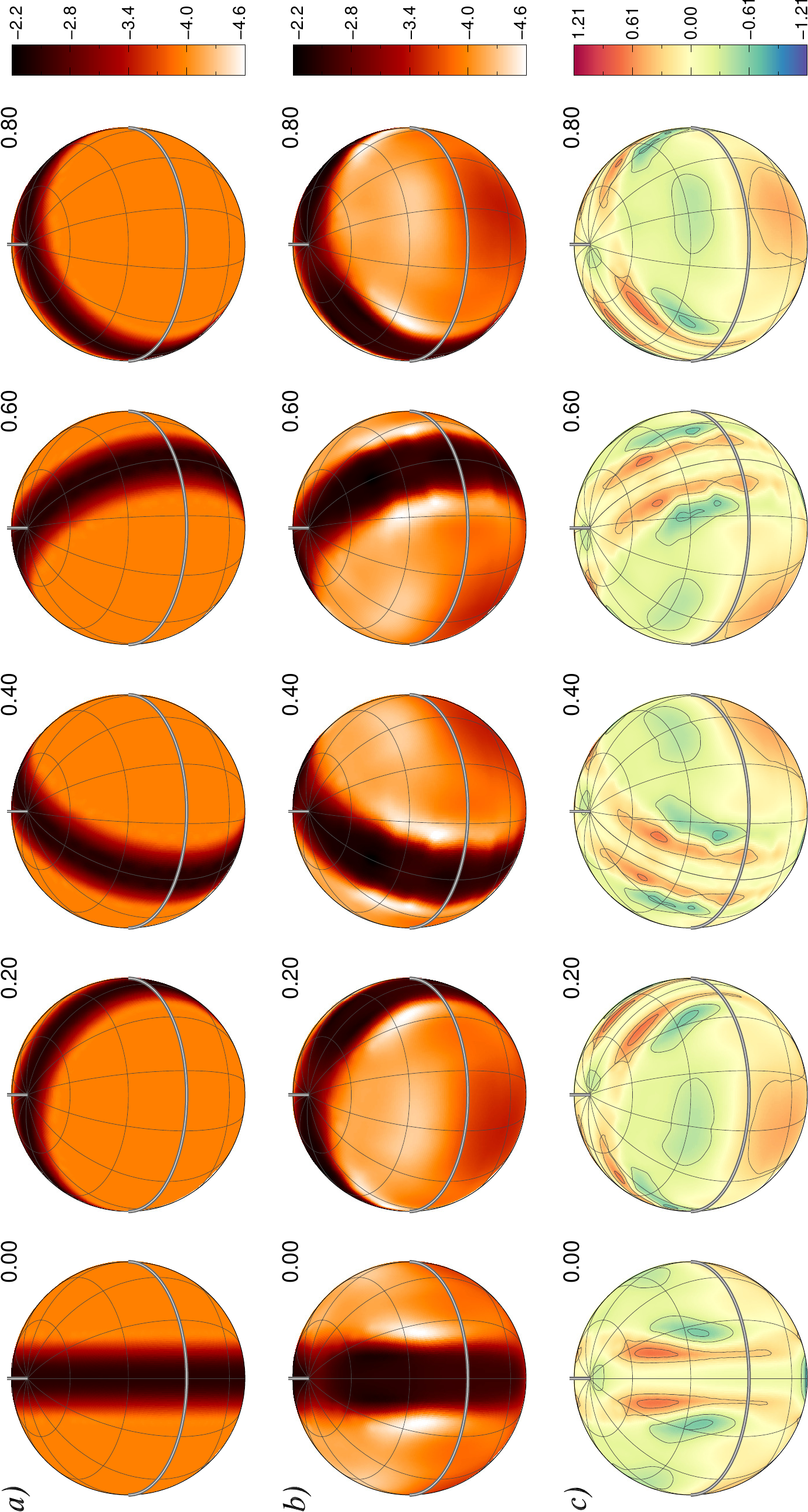}
\caption{Same as Fig.~\ref{fig:map2} but ignoring effect of local abundance variation on the model atmosphere structure. 
}
\label{fig:map9}
\end{figure*}

In addition to the numerical error assessment reported in Table~\ref{tbl:tests}, we present a comparison between the true and recovered abundance maps in a series of spherical plots (Figs.~\ref{fig:map1}-\ref{fig:map9}). Each such figure is composed of three rows of panels showing the input abundance map, reconstructed map and the corresponding difference map after subtracting the mean offset. The true and reconstructed maps are rendered using the same linear colour scale, without applying any cuts to the maximum and minimum values. The difference map is shown using a different colour table scaled to the same abundance range -- this allows one to judge the significance of error map structures in comparison to the input and recovered abundance distributions. Patterns in the difference maps are further highlighted with contours plotted every 0.3~dex.

\subsection{Inversion results}

\subsubsection{Multiple circular spots}
\label{refs}

We started by assessing accuracy of the DI reconstruction of the multiple circular spot abundance map under the optimal conditions, taking into account the stellar magnetic field. The true and recovered maps for this test are shown in Fig.~\ref{fig:map1}. As evident from this figure and Table~\ref{tbl:tests}, very accurate inversion results are achieved. There is no significant mean offset between the true and recovered maps. The average reconstruction errors are 0.07--0.09~dex while the maximum errors are 0.11--0.12~dex. The largest errors are found for the region occupied by the sub-equatorial overabundance spot. This structure is inferred to have a somewhat larger extension and a lower contrast than in the input map. A careful inspection of the difference map shows a qualitatively similar but much less pronounced smoothing of higher latitude spots. These reconstruction artefacts are, however, entirely negligible compared to the 1.5~dex contrast of the input non-uniform Fe distribution.

\subsubsection{Overabundance ring}
\label{refc}

In the second test we assessed the ability of our DI code to recover a ring-like overabundance structure. Similar to the previous test, magnetic field is fully accounted for in spectrum synthesis and no systematic errors are introduced in the DI reconstruction. The results of this experiment are illustrated in Fig.~\ref{fig:map2}. Again, we find a successful recovery of the true overabundance structure, with average reconstruction errors of 0.06~dex, maximum errors of 0.08~dex, and no systematic offset between the input and recovered maps. As before, the largest discrepancy between the maps is found for the lowest visible latitude belts ($\le$\,$-30\degr$), which contribute negligibly to the disk-integrated spectra and, therefore, are poorly constrained by the inversion.

\subsubsection{Inversions based on a few spectral lines}
\label{2lines}

Among many chemical elements studied by the DI technique in Ap/Bp stars only a few species (Si, Cr, Fe) present a large number of useful diagnostic lines in the spectra of moderately and fast-rotating ($v_{\rm e}\sin i=30$--100~\kms) stars. The majority of elements can be studied using only one or two lines because other lines are unsuitable due to blending or absent altogether in the optical wavelength range typically covered by spectroscopic observations. In this context, we are interested to ascertain the usefulness of a one or two-line DI inversion compared to using five or more lines with diverse parameters. To this end, we tested reconstruction of the 4-spot Fe abundance distribution using the \ion{Fe}{ii} 6147.7 and 6149.2~\AA\ lines (originating from the same multiplet and having the same strength) instead of the 7 \ion{Fe}{i} lines (having different strengths and excitation potentials) employed above. The purpose of this experiment is not to analyse a degradation of the DI images with increased observational noise. Therefore, we have increased the signal-to-noise ratio of simulated observational data to compensate for the reduction of the number of wavelength points.

The results of the inversion using the two \ion{Fe}{ii} lines are presented in Fig.~\ref{fig:map3}. One can see, based on this figure and information in Table~\ref{tbl:tests}, that the reconstruction errors slightly increase compared to the inversion documented in Sect.~\ref{refs}. In particular, the Fe concentration is underestimated at the centre of the $-30\degr$ latitude spot. The spot at $+30\degr$ is shifted to lower latitudes by 5--10\degr. Nevertheless, the average errors are still only 0.10--0.11~dex and the maximum errors are 0.13--0.15~dex, which is small compared to the 1.5~dex abundance contrast in the input map.

We have carried out a further test of multi-element DI mapping using a single blended line by adding the \ion{Cr}{ii} 6146.2~\AA\ line to the list of the two \ion{Fe}{ii} lines. The same ring-like element distribution as studied in Sect.~\ref{refc} was adopted for Cr. This experiment represents a challenging test of the simultaneous recovery of two abundance maps since for the adopted $v_{\rm e}\sin i$ the \ion{Cr}{ii} line is an unresolved, weaker component of the blend with \ion{Fe}{ii} 6147.7~\AA.

Figure~\ref{fig:map4} shows the Cr and Fe abundance maps reconstructed simultaneously by {\sc Invers10}. The quality of the Fe map reconstruction (average error 0.11--0.13~dex, maximum error 0.15--0.16~dex) is comparable to the inversion using the \ion{Fe}{ii} lines alone. At the same time, the errors of the Cr ring-like map recovery (average error 0.08, maximum error 0.11) are almost the same as achieved in Sect.~\ref{refc} for the inversion based on 7 \ion{Fe}{i} lines.

Based on these tests, we conclude that there is no appreciable gain, apart from decreasing observational noise, in using many spectral lines for chemical spot DI of Ap/Bp stars. Abundance mapping with a single line, even a blended one, is not significantly inferior to an inversion based on five or more lines provided that different blend contributors are identified correctly and modelled accurately.

\subsubsection{Ignoring a moderately strong magnetic field}
\label{magtest}

As mentioned above, a significant fraction of previous abundance DI studies of Ap/Bp stars have ignored the magnetic field in chemical spot mapping. This potentially leads to systematic errors and artefacts in the reconstructed maps. The goal of this section is to quantify these effects and compare possible errors with typical horizontal abundance contrasts.

The influence of the magnetic field on Stokes $I$ line profiles and therefore its possible impact on chemical abundance DI strongly depends on the magnitude of the field and on the sensitivity of a given spectral line to the Zeeman effect. A line with a small or null effective Land\'e can be reliably studied with non-magnetic spectral modelling methods even in strong-field targets. On the other hand, typical spectral lines with effective Land\'e factors 0.9--1.5, such as the group of 7 \ion{Fe}{i} lines studied here, exhibit a noticeable magnetic local profile modification for the field strength of $\ga$\,2~kG. However, these profile shape distortions are not observable for moderately or rapidly rotating Ap/Bp stars typically targeted by DI. Instead, the primary manifestation of the Zeeman effect is a modification of the local equivalent widths of spectral lines through the magnetic intensification effect, which changes across the stellar surface according to the local field strength and orientation. This magnetic line strengthening might be misinterpreted as chemical inhomogeneities by a non-magnetic DI code.

It is essential to adopt a representative magnetic field strength and geometry for an assessment of systematic errors of an inversion ignoring magnetic field. To this end, the 2.5~kG polar field strength of the dipolar topology assumed in our experiments corresponds to the average field strength of a volume limited sample of nearby Ap/Bp stars \citep{power:2008}. On the other hand, the average dipolar field strength for a sample of non-magnetic abundance DI targets discussed below (see Sect.~\ref{disc}) is 2.0~kG.

To study the effects of neglecting magnetic field in chemical abundance mapping we have repeated the inversions discussed in Sects.~\ref{refs} and \ref{refc} using the same simulated data but assuming zero field instead of the correct 2.5~kG dipolar field topology. Furthermore, since the magnetic intensification of individual spectral lines leads to a line-to-line scatter of the mean equivalent width, which in real DI applications is difficult to separate from other systematic errors (for instance, erroneous oscillator strengths), we have enabled an automatic oscillator strength correction for all \ion{Fe}{i} lines from the list in Table~\ref{tbl:lines} except \ion{Fe}{i} 5005.7 and 5006.1~\AA.

The results of abundance inversions ignoring the magnetic field are illustrated in Fig.~\ref{fig:map5} for the multiple circular spot distribution and in Fig.~\ref{fig:map6} for the chemical overabundance ring. In both cases we detect a systematic offset on the order of 0.3~dex between the true and reconstructed maps. This offset is subtracted before further analysis of the difference map. Not surprisingly, Figs.~\ref{fig:map5} and \ref{fig:map6} show that inversion errors increase when the magnetic field is ignored in abundance mapping. The contrast of reconstructed abundance maps is somewhat larger than for the case of correct magnetic field treatment. The maps also show a systematic latitude dependence of reconstruction errors: the relative abundance is underestimated for higher latitudes but overestimated for lower latitudes. The fit to simulated observations is also not as good as before, with the standard deviation at convergence larger by some 20\% in the ring-structure inversion compared to the test in Sect.~\ref{refc}. These issues are reflected in the average errors of 0.19--0.22~dex and maximum errors of 0.28--0.33~dex. These errors are about a factor of two larger than those found for the optimal inversions in Sects.~\ref{refs} and \ref{refc} but still not overwhelming compared to the 1.5~dex input abundance contrast and certainly do not preclude a correct identification of the main chemical overabundance features at the stellar surface.

We have carried out an additional inversion to assess a spurious component of the abundance DI maps resulting from neglecting magnetic field. In this test simulated observations were generated for a uniform Fe distribution with $\log N_{\rm Fe}/N_{\rm tot}=-4.0$ and the same 2.5~kG dipolar field as before. Then an abundance inversion was carried out assuming zero magnetic field. In this case all structures in the resulting chemical spot map, shown in Fig.~\ref{fig:map7}, are spurious. Quantitative assessment indicates an overestimation of the mean abundance by 0.34~dex, the average errors of 0.13~dex, and the maximum errors of 0.18--0.19~dex. These errors are smaller than in the inversions with inhomogeneous surface abundance distributions. The structures in the difference map in Fig.~\ref{fig:map7} are also far less conspicuous than in Figs.~\ref{fig:map5}c and \ref{fig:map6}c. This suggests that the primary effect of an unaccounted magnetic field on abundance DI is to distort a real chemical abundance distribution rather than to produce a completely spurious abundance structure. Thus, the non-magnetic abundance inversion errors include two contributions: spurious abundance structures mimicking the effect of local magnetic line intensification ($\sim0.1$~dex for the 2.5~kG dipolar field considered here) and abundance map distortions due to an incorrect recovery of chemical inhomogeneities present on the stellar surface. The latter contribution scales with the surface element distribution contrast, reaching $\sim$\,0.2~dex for the 1.5~dex spot to photosphere abundance contrast considered in our experiments.

\subsubsection{Effect of local atmospheric structure}
\label{i13test}

Apart from the Si and Fe abundance inversions discussed by \citet{lehmann:2007} for HR\,7224 and by \citet{kochukhov:2012} for $\alpha^2$~CVn, respectively, and mapping of He spots in a few hot CP stars \citep{yakunin:2015,oksala:2015}, no systematic effort was made by previous DI studies to achieve a self-consistency between chemical spot inversions and lateral variation of the atmospheric structure and continuum brightness. Since the metal overabundance spots are brighter in the optical than the surrounding atmosphere \citep[e.g.][]{luftinger:2010a,krticka:2015}, an abundance DI based on a single, average stellar model atmosphere may overestimate the contrast of surface abundance distribution and lead to other artefacts. However, the studies by \citet{lehmann:2007} and \citet{kochukhov:2012} did not reveal a major influence of this problem on DI maps.

In this section we further explore and characterise the systematic DI errors and artefacts related to neglecting an impact of chemical abundance spots on the local atmospheric structure. To this end, we have generated simulated observational data with the same characteristics and for the same magnetic field and abundance geometries as discussed above using forward calculations with the {\sc Invers13} code and a grid of {\sc LLmodels} atmospheres covering the full range of Fe abundances present in the input maps. The local line and continuum spectra were thus calculated fully self-consistently with the adopted local Fe abundance. We then analysed these simulated observations with the {\sc Invers10} code, using a single model atmosphere calculated for $\log N_{\rm Fe}/N_{\rm tot} = -4.0$. The resulting comparison between the true and recovered abundance maps is presented in Fig.~\ref{fig:map8} and Fig.~\ref{fig:map9} for the multiple spots and overabundance ring surface distribution scenarios respectively.

According to Table~\ref{tbl:tests}, the average abundance reconstruction errors stemming from an inconsistent treatment of chemical spots and local atmospheric structure are about 0.2~dex while the maximum errors reach 0.3~dex. On the other hand, no significant mean abundance offset is detected in the reconstructed maps. As evident from Figs.~\ref{fig:map8} and \ref{fig:map9}, the spot maps obtained with a single model atmosphere exhibit a somewhat higher abundance contrast, with spurious underabundance zones found close to or in between the true overabundance features. In addition, one can find a couple of small areas in the difference map for the multiple circular spot test case (Fig.~\ref{fig:map8}c) where the Fe abundance appears to be significantly overestimated. These artefacts correspond to a latitudinal smearing by 10--20\degr\ of two real overabundance spots.

In general, results of the experiments presented in this section lead to the conclusion that the atmospheric structure variation due to the 1.5~dex Fe abundance contrast adopted in our tests has a marginal impact on the classical abundance DI. On the other hand, abundance anomalies of elements other than Fe are less important for the atmospheric structure of A and B-type stars \citep{khan:2007} and hence their impact on DI maps is even smaller than discussed here. These considerations suggest that a self-consistent treatment of metal overabundance spots and local atmospheric structure is not a critical factor determining reliability of CP star abundance DI maps.

\section{Discussion}
\label{disc}

\begin{table*}[!th]
%\centering
\caption{Doppler imaging studies of chemically peculiar stars
\label{tbl:distudies}}
{\small
\begin{tabular}{llrllp{4cm}l}
\hline\hline
HD & Name & $T_{\rm eff}$~~ & Magnetic field & DI method & Chemical maps & Reference \\
\hline
\multicolumn{7}{c}{\textit{Ap/Bp stars}}\\
3980 & $\xi$~Phe & 8300 & $6.9^1$ & DI & Li, O, Si, Ca, Cr, Mn, Fe, La, Ce, Pr, Nd, Eu, Gd & \citet{nesvacil:2012} \\
11503 & $\gamma^2$~Ari & 10250 & $3.0^1$ & DI & Si & \citet{hatzes:1989}\\ 
15089 & $\iota$~Cas & 8350 & $2.0^1$ & DI & Cr & \citet{kuschnig:1998a}\\ 
18296 & 21~Per & 9350 & $1.6^1$ & DI & Si, Ti, Cr, Mn, Fe & \citet{wehlau:1991}\\ 
19832 & 56~Ari & 12800 & $1.3^1$ & DI & Si & \citet{ryabchikova:2003b}\\ 
24712 & HR\,1217 & 7250 & $3.6^1$ & MDI & Mg, Ca, Sc, Ti, Cr, Fe, Co, Ni, Y, La, Ce, Pr, Nd, Gd, Tb, Dy & \citet{luftinger:2010}\\ 
 & & & & MDI & Na, Fe, Nd &\citet{rusomarov:2015}\\
32633 & HZ~Aur & 12800 & $17^2$ & MDI & Mg, Si, Ti, Cr, Fe, Ni, Nd & \citet{silvester:2015} \\
37479 & $\sigma$~Ori~E & 23000 & $7.5^1$ & MDI & He, C, Si, Fe & \citet{oksala:2015}\\
37776 & V901~Ori & 22000 & $30^2$ & MDI & He, O, Si, Fe & \citet{khokhlova:2000} \\
 & & & & MDI & He & \citet{kochukhov:2011a}\\
40312 & $\theta$~Aur & 10400 & 1.3$^1$ & DI & Si, Cr, Fe & \citet{rice:1990}\\ 
 & & &  & DI & Si, Cr & \citet{hatzes:1991} \\
 & & & & DI & He, Mg, Si, Ti, Cr, Mn, Fe & \citet{kuschnig:1998b} \\
 & & &  & DI & O & \citet{rice:2004}\\
50773 & & 8300 & $2.0^1$ & DI & Mg, Si, Ca, Ti, Cr, Fe, Ni, Y, Cu & \citet{luftinger:2010a} \\
55522 & HR\,2718 & 17400 & $2.6^3$ & DI & He, Si & \citet{briquet:2004}\\ 
65339 & 53~Cam & 8400 & $26^2$ & MDI & Si, Ca, Ti, Fe, Nd & \citet{kochukhov:2004d}\\ 
72106 & & 11000 & $1.2^1$ & MDI & Si, Ti, Cr, Fe & \citet{folsom:2008}\\
75049 & & 10250 & $36^1$ & MDI & Si, Cr, Fe, Nd & \citet{kochukhov:2015} \\
79158 & 36~Lyn & 13300 & 3.4$^1$ & DI & Fe & \citet{wade:2006a} \\ 
83368 & HR\,3831 & 7650 & 2.5$^1$ & DI & Li, C, O, Na, Mg, Si, Ca, Ti, Cr, Mn, Fe, Co, Ba, Y, Pr, Nd, Eu & \citet{kochukhov:2004e}\\
105382 & HR\,4618 & 17400 & $2.3^1$ & DI & He, Si & \citet{briquet:2004}\\ 
108662 & 17~Com & 10300 & $3.3^1$ & DI & Cr, Fe & \citet{rice:1994} \\ 
112185 & $\varepsilon$~UMa & 9000 & 0.4$^1$ & DI & O & \citet{rice:1997}\\
 & & & & DI & Ca, Cr, Fe, Mg, Mn, Ti, Sr & \citet{lueftinger:2003} \\
112413 & $\alpha^2$~CVn & 11600 & $5.5^2$ & MDI & O, Si, Cl, Ti, Cr, Fe, Pr, Nd, Eu & \citet{silvester:2014a} \\ 
120198 & 84~UMa & 10450 & $1.6^1$ & DI & Cr, Fe &\citet{rice:1994a}\\ 
124224 & CU\,Vir & 12750 & $3.0^1$ & DI & He, Mg, Si, Cr, Fe & \citet{kuschnig:1999}\\
 & & & & MDI & Si, Fe & \citet{kochukhov:2014} \\
125248 & CS\,Vir & 9850 & $11^2$ & MDI & Ti, Cr, Fe, Ce, Nd, Gd & \citet{rusomarov:2016} \\ 
125823 & a~Cen & 19000 & $1.6^1$ & DI & He & \citet{bohlender:2010}\\ 
131120 & HR\,5543 & 18250 & $0.7^3$ & DI & He, Si & \citet{briquet:2004}\\
138769 & HR\,5781 & 17500 & $0.9^3$ & DI & Si & \citet{briquet:2004}\\ 
140728 & BP~Boo & 10500 & $1.4^3$ & DI & Si & \citet{hatzes:1990}\\ 
148112 & $\omega$~Her & 9350 & $1.0^1$ & DI & Cr & \citet{hatzes:1991a}\\  
151525 & 45~Her & 9400 & $0.5^1$ & DI & Cr & \citet{hatzes:1991a}\\ 
170000 & $\varphi$~Dra & 12500 & $1.8^1$ & DI & He, Si, Ti, Cr, Fe & \citet{kuschnig:1998b} \\ 
177410 & HR\,7224 & 14500 & $1.3^3$ & DI & Si, Fe & \citet{lehmann:2007}\\ 
182180 & HR\,7355 & 17500 & $11.6^1$ & MDI & He & \citet{rivinius:2013} \\
184927 & V1671~Cyg & 22000 & $9^2$ & MDI & He, O & \citet{yakunin:2015} \\ 
219749 & ET~And & 11500 & $1.4^3$ & DI & He, Si & \citet{piskunov:1994} \\ 
& & & & DI & He, Mg, Si, Ti, Cr, Fe & \citet{kuschnig:1998b} \\
220825 & $\kappa$~Psc & 9250 & $2.0^1$ & DI & Cr & \citet{ryabchikova:1996}\\ 
\hline
\multicolumn{7}{c}{\textit{HgMn stars}}\\
358 & $\alpha$~And~A & 13800 & non-magnetic & DI & Hg & \citet{kochukhov:2007b}\\
11753 & $\varphi$~Phe & 10500 & non-magnetic & DI & Ti, Cr, Sr, Y & \citet{makaganiuk:2012}\\
 & & & & DI & Ti, Cr, Sr, Y & \citet{korhonen:2013}\\
32964A & 66~Eri~A & 11100 & non-magnetic & DI & Ti, Sr, Y, Ba & \citet{makaganiuk:2011}\\
34364A & AR~Aur~A & 10950 & non-magnetic & DI & Ti, Fe, Y & \citet{hubrig:2010}\\
\hline
\end{tabular}
\tablefoot{Magnetic field strength parameter: (1) -- polar strength of a dipolar field topology, (2) -- maximum local field modulus for a non-dipolar field topology, (3) -- maximum absolute mean longitudinal field multiplied by 3. All field strengths are given in kG.}
}
\end{table*}

In preceding sections we have examined an impact of several systematic error sources, such as neglecting the magnetic field, on the Doppler imaging reconstruction of chemical spot maps in early-type stars. What are the implications of these results for the interpretation of published chemical abundance distributions? To address this question, we have complied a summary of all chemical abundance DI studies which were based on reasonably modern, high-quality observations (i.e. high signal-to-noise ratio spectra recorded with solid-state detectors) published between 1989 and 2016. This summary, presented in Table~\ref{tbl:distudies}, lists 36 magnetic Ap/Bp stars investigated in 39 individual DI and MDI studies. A few DI publications superseded by later studies of the same chemical elements in the same stars were omitted. 

For each star Table~\ref{tbl:distudies} provides an estimate of the peak surface magnetic field strength, $B_{\rm max}$. This parameter was taken from the MDI magnetic field maps whenever available or adopted according to the dipolar field strength reported in the literature. For a few stars the existing magnetic measurements are insufficient to constrain even the simplest dipolar field model. In these cases we adopted $3|\langle B_{\rm z}\rangle|_{\rm max}$ as an approximation of $B_{\rm max}$. 

For completeness Table~\ref{tbl:distudies} also includes information on four HgMn stars investigated with abundance DI. These stars are qualitatively different in their spectral appearance and characteristics of non-uniform element distributions (low-contrast and evolving) from magnetic Ap/Bp stars. Although extremely weak fields of the type recently found in Vega and Sirius \citep{petit:2010,petit:2011} cannot be excluded for HgMn stars, these objects certainly lack the organised global fields typical of classical magnetic CP stars \citep{makaganiuk:2011a,kochukhov:2013a}.

\begin{figure}[!t]
\centering
\fifps{7.5cm}{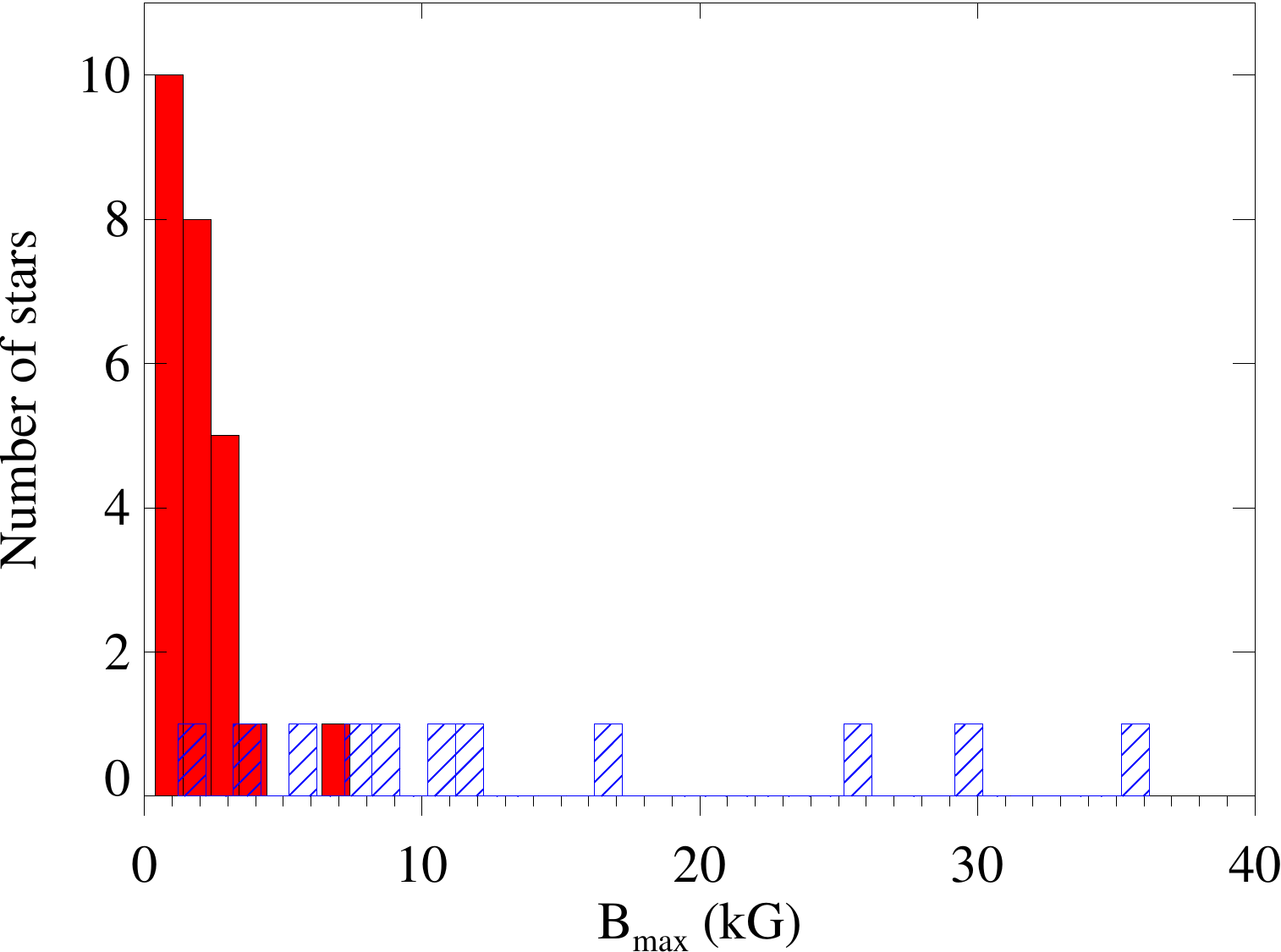}
\caption{Distribution of peak surface magnetic field strengths for DI (filled histogram bins) and MDI (hatched histogram bins) targets.}
\label{fig:stat}
\end{figure}

The distribution of peak surface magnetic field strength for 36 magnetic Ap/Bp stars with available chemical spot maps is shown in Fig.~\ref{fig:stat}. The MDI targets span a wide range of $B_{\rm max}$ from $\sim$\,1 to $>$\,30~kG. On the other hand, the stars studied with non-magnetic DI techniques are concentrated at lower field strengths. Their average $B_{\rm max}$ is equal to 2.0~kG and exceeds 3.0~kG for only 3 out of 25 stars. HD\,3980 with its 6.9~kG dipolar field \citep{nesvacil:2012} is a notable outlier. However, the field strength and topology of this star were constrained using only 5 mean longitudinal magnetic field measurements and thus are by no means definitive. The presence of a strong field in HD\,3980 needs to be confirmed with high-resolution spectropolarimetric observations.

Given the results of our numerical tests in Sect.~\ref{magtest}, which demonstrated only minor effects of ignoring a 2.5~kG dipolar field, there is no evidence that a systematic neglect of the magnetic field is detrimental for the quality of chemical maps for the majority of DI targets with fields of a few kG or less. In fact, only the non-magnetic DI analysis of HD\,3980 by \citet{nesvacil:2012} may potentially suffer from non-negligible artefacts for some elements provided that the strong magnetic field of this star is real. One can ascertain though that the majority of chemical abundance maps of HD\,3980 cannot be dominated by spurious effects of an unaccounted Zeeman line broadening and intensification since the derived element distributions are diverse and do not follow a clear common pattern. In any case, quantifying the probable inversion errors requires a separate element-by-element assessment that takes into account the magnetic field response of specific spectral lines employed for the DI mapping of this star.

\section{Summary and conclusions}
\label{summary}

In this study we presented a series of numerical experiments addressing the question of reliability of chemical abundance maps derived for early-type magnetic stars with the Doppler imaging technique. In particular, we focus on assessing the systematic errors of DI reconstructions coming from common simplifying approximations, such as neglect of a magnetic field in radiative transfer and ignoring horizontal variation of the stellar atmosphere structure. These issues were investigated for a set of atmospheric parameters corresponding to a cool Ap star. At the same time, none of the studied systematic effects is expected to have a significant temperature dependence, suggesting that our conclusions are also valid for hotter Ap/Bp stars.

The main results of our investigation can be summarised as follows.
\begin{itemize}
\item Our DI code achieves an average accuracy of 0.06--0.09~dex and maximum errors of 0.12~dex for the chemical map reconstruction which include magnetic field effects in spectrum synthesis and when using about half a dozen of individual spectral lines.
\item DI reconstruction is equally successful for a complex element distribution comprised of multiple circular element overabundance spots and for a narrow overabundance ring located at the magnetic equator of a dipolar field.
\item A reduction of the line list to 1--2 spectral features, including blended one, leads to only a marginal increase of the average inversion errors to about 0.1~dex and maximum errors to $\sim$\,0.15~dex. Therefore, apart from counteracting random observational noise, abundance DI inversions do not gain significantly from modelling a large number of spectral lines.
\item Ignoring a moderately strong, 2.5~kG dipolar magnetic field in abundance DI introduces a mean offset of about 0.3~dex in the recovered chemical maps. The average relative reconstruction errors increase to $\sim$\,0.2~dex while the maximum relative errors reach $\sim$\,0.3~dex. These errors correspond to distortions of real surface abundance inhomogeneities. The spurious component of abundance maps (i.e. Zeeman broadening and intensification misinterpreted as abundance variations) does not exceed $\sim$\,0.15~dex on average.
\item Ignoring local atmospheric structure variations in the areas of element overabundance leads to average reconstruction errors of $\sim$\,0.2~dex and maximum errors of $\sim$\,0.3~dex. These numbers correspond to the effect of overabundance spots of the element (iron) providing the most important contribution to the atmospheric opacity. The artefacts resulting from using a single average model atmosphere for DI inversions correspond to the distortion of real chemical inhomogeneities; no spurious spots are produced.
\item Reviewing results of several dozen non-magnetic DI studies in the light of our findings, we conclude that the absolute majority of these investigations have targeted stars with sufficiently weak fields and therefore could not have been adversely affected by the neglect of these fields in abundance mapping. In general, the systematic DI reconstruction errors (up to 0.2--0.3~dex) inferred in this paper are small  or entirely negligible compared to horizontal element abundance contrasts of 2--5~dex typically derived with DI for Ap/Bp stars.
\end{itemize}

\begin{acknowledgements}
The author thanks Drs. T. L\"uftinger and J. Silvester for comments on this paper. This research is supported by the Knut and Alice Wallenberg Foundation, the Swedish Research Council, and the Swedish National Space Board.
\end{acknowledgements}

%\bibliographystyle{aa}
%\bibliography{astro_papers}

\end{document}